\documentclass[aps,prd,nofootinbib,onecolumn,superscriptaddress,preprintnumbers,longbibliography]{revtex4-2}

\usepackage{placeins}
\usepackage{amsmath,amssymb,mathtools,bm}
\usepackage{comment}
\usepackage{gensymb}
\usepackage{physics}
\usepackage{cancel}
\usepackage[normalem]{ulem} 

\usepackage{tablefootnote}
\usepackage{graphicx, color, hepunits}
\usepackage[dvipsnames]{xcolor}
\usepackage{float}
\usepackage{filecontents}
\usepackage{multirow}
\usepackage{slashed}
\usepackage{pifont}
\usepackage[colorlinks=true 
,urlcolor=purple 
,anchorcolor=blue
,citecolor=blue 
,filecolor=magenta  
,linkcolor=blue 
,menucolor=blue
,linktocpage=true
,pdfproducer=medialab
]{hyperref}

\usepackage[utf8]{inputenc}
\usepackage[english]{babel}

\restylefloat{table}

\begin{document}
\preprint{CERN-TH-2026-087}

\title{How well can the QCD axion hide?} 

\author{Sung Mook Lee}
\email{sungmook.lee@cern.ch}
\affiliation{Theoretical Physics Department, CERN, CH-1211 Gen\`eve 23, Switzerland}

\author{Maria Ramos}
\email{maria.ramos@cern.ch}
\affiliation{Theoretical Physics Department, CERN, CH-1211 Gen\`eve 23, Switzerland}

\author{Fuensanta Vilches}
\email{fuenvilches@ugr.es}
\affiliation{Departamento de Física Teórica y del Cosmos, Universidad de Granada, Campus de Fuentenueva, E-18071 Granada, Spain}

\begin{abstract}

Motivated UV frameworks generically predict the existence of multiple axion fields.
Their interplay gives rise to novel collective phenomena -- including level crossings and the formation of string bundles -- which modify the predicted mass and couplings of the QCD axion as a solution to both the strong CP problem and the observed dark matter abundance.
Among these effects, the domain wall number is determined by the full anomaly structure of the theory: in the single axion case, the absence of long-lived domain walls imposes $E/N \geq 8/3$ as a theoretical bound on the QCD axion photon coupling, assuming the global structure of the Standard Model gauge group is minimal.
We show that this bound can be relaxed in the multi-axion framework.
Combined with the fact that the QCD axion can become a subdominant dark matter component, this might render multi-axion scenarios experimentally challenging.
Nevertheless, a careful analysis of the parameter space reveals that in most regions where the QCD axion evades detection, an axion-like particle remains visible to next-generation experiments. 
When all signals fall below future projections, we identify the most promising regions of parameter space to probe in an illustrative two-axion setup.

\end{abstract}

\maketitle

\section{Introduction}
\label{sec:intro}

The QCD axion is one of the most compelling new physics candidates which can solve two prominent problems at once.
It solves the strong CP problem by dynamically absorbing the QCD $\theta$-angle, while explaining the dark matter relic abundance in a predictive region of the parameter space.

Non-perturbative QCD effects generate an axion mass $ m_{a} \sim \Lambda_{\rm QCD}^{2} / f_{a} $, that is fully determined by its coupling to gluons, $1/ f_{a} $, and the QCD confinement scale, $ \Lambda_{\rm QCD}$.
In a post-inflationary cosmology, this relation fixes $m_a 
\sim 10^{-4}\,\text{eV}$~\cite{Sikivie:2006ni,Marsh_2016} if the QCD axion accounts for the totality of observed dark matter.
In this scenario, the axion number density is set by string network dynamics and misalignment, assuming domain walls (DWs) do not dominate the energy density of the universe.
This constrains the axion potential to have a single minimum ($N_{\rm DW}=1$), so each string has exactly one DW attached. The tension of this wall pulls the strings towards each other, causing the network to annihilate and avoiding the DW problem.

While this mechanism constrains the gluonic coupling of the axion, most experimental advances are focused on its coupling to photons,
\begin{equation}
    g_{a\gamma} =\frac{\alpha_{\rm em}}{2\pi f_a} \left( \frac{E}{N} - 1.92 \right) \,,
    \label{eq:gagg}
\end{equation}
where $ \alpha_{\rm em} $ is the fine structure constant.
The coupling in Eq.~\eqref{eq:gagg} depends on the UV model that completes the axion effective field theory (EFT) via the $E/N$ factor, the ratio between the electromagnetic and color anomalies. 
The second contribution to this coupling is an IR effect generated by the mixing of the axion with the neutral pion.
Even though this contribution breaks coupling quantization in the minimal model, it is computable.
Therefore, a potential measurement could still provide valuable information about the UV content of the axion EFT.

Current targets of axion experiments are mainly set by specific UV models.
For example, 
$E/N=0$ can be obtained in KSVZ models~\cite{Kim:1979if,Shifman:1979if}, 
while $ E/N=8/3 $ is a prediction of DFSZ 
models~\cite{Zhitnitsky:1980tq,Dine:1981rt} as well as of minimal Grand Unified Theories (GUTs)~\cite{Raffelt:2006rj,Giudice:2012zp,Agrawal:2022lsp}.
However, under different UV charge assignments~\cite{DiLuzio:2020wdo}, 
$E/N\approx 2$ can also be obtained, which largely cancels the IR contribution to the axion photon coupling. This shows that $g_{a\gamma}$ is not theoretically bounded, and experimental targets based on benchmark models may miss a significant region of axion parameter space.

Interestingly, recent works~\cite{Reece:2023iqn,Choi:2023pdp,Cordova:2023her} have shown that the global structure of the Standard Model (SM) gauge group can place constraints on $E/N$, independently of the UV completion. Such constraints depend on the choice for the discrete quotient group \cite{Tong:2017oea}:
\begin{equation}
    G_{\rm SM} = \frac{SU(3)_{C} \times SU(2)_{L} \times U(1)_Y}{\mathbb Z_p } \, ,
    \label{eq:minG}
\end{equation}
with $p \in \{ 1,2,3,6 \}$. Although the local dynamics is unchanged by this choice, different values of $p$ define globally distinct theories, e.g. differing in the allowed number of representations of electrically charged particles.
In particular, the case $p=6$ admits the smallest set of these representations, and we therefore refer to it as the minimal case.
We also note that embedding the SM into minimal GUT theories based on 
$SU(5)$ or $SO(10)$ selects $ p = 6$ \cite{Baez:2009dj}.

Depending on the value of $ p $, the axion couplings to gauge bosons obey correlated quantization conditions.
In particular, if $G_{\rm SM}$ is minimal ($p=6$) and $N_{\rm DW}=1$, it has been shown  that
\begin{equation}
    g_{a\gamma}  
    \geq \frac{\alpha_{\rm em}}{2\pi f_a} (8/3-1.92)\,, 
    \label{eq:mincoup}
\end{equation}
for a post-inflationary axion \cite{Reece:2023iqn,Choi:2023pdp}.

In this work, we investigate whether these predictions remain robust in multi-axion frameworks.
Not only is minimality an ad hoc assumption of the standard axion mechanism, but the most compelling UV completions of the QCD axion generically predict additional axion fields.
This is the case of string theory, one of the most promising avenues to address the axion quality problem, from which one expects an~\textit{axiverse}~\cite{Arvanitaki:2009fg,Demirtas:2021gsq}.
Other examples include more general higher-dimensional theories~\cite{Dienes:2011ja, Dienes:2011sa,deGiorgi:2024elx}; models where QCD is embedded into a product gauge group, typically requiring more than one axion to shield the contributions from additional $\theta$-angles~\cite{Agrawal:2017ksf,Csaki:2019vte,Fuentes-Martin:2019bue}; or GUTs with a non-minimal spectrum of pseudo-Goldstone bosons (charged under the GUT group but SM singlets) which can mix with the QCD axion~\cite{Agrawal:2022lsp}.

As a first step towards extending the QCD axion paradigm, we consider a two-axion model with two confining scales, one induced by QCD and the other by a dark gauge group.
While the dark axion largely decouples from the strong CP problem solution across the parameter space, 
we show that it can significantly alter the predictions of the single QCD axion model, modifying the minimal photon coupling in Eq.~\eqref{eq:mincoup} and the mass at which the observed relic abundance is explained.
The key point is that cosmology is sensitive to the UV content of the theory: even if additional axion-like fields are heavier than the QCD axion and decouple from IR physics, the topology of the vacuum manifold -- and hence the properties of cosmological defects -- retains information about the field content. The same argument holds for particles which are lighter than the QCD axion, but still impact cosmology by mixing sizably with the QCD axion at high temperatures.

We note that some aspects of multi-axion phenomenology have been discussed previously, including their impact on the strong CP problem~\cite{Agrawal:2017ksf,Higaki:2016jjh,Hu:2020cga}, the misalignment mechanism~\cite{Kitajima:2014xla,Dunsky:2025sgz}, the evolution of topological defects~\cite{Choi:1985iv,Higaki:2016jjh,Lee:2024toz}, inflation~\cite{Kim:2004rp,Dimopoulos:2005ac,Choi:2014rja}, and generalized symmetries~\cite{Hong:2025qbw}.
However, a systematic study of their combined effects has not been presented. 
This is critical to assess how non-minimal (more realistic) axion frameworks distort the QCD axion signal regions, and guide the future experimental programme.

Our work is structured as follows. In Sec.~\ref{sec:setup}, we present the model and discuss the physical couplings of the two axions, including their quantization relations. In Sec.~\ref{sec:DWnumber}, we obtain the generalized DW constraint and in Sec.~\ref{sec:minc} we apply this constraint to find the minimal QCD axion coupling to photons. In Sec.~\ref{sec:dm}, we compute the dark matter abundances and identify the parameter space in which the two axions comprise all the observed relic density. In Sec.~\ref{sec:pheno}, we obtain constraints on this parameter space, assuming the minimal coupling scenario. We conclude in Sec.~\ref{sec:conclusions}.

\section{Two axion setup}\label{sec:setup}

We consider a model of two axions $a_{1,2}$ associated to two $U(1)$ Peccei-Quinn (PQ) symmetries, and with periodicities given by~
\begin{equation}
    a_{1,2} \sim a_{1,2} + 2 \pi f_{1,2}\,.
    \label{eq:shifts}
\end{equation}
These two $U(1)_{1,2}$ symmetries are spontaneously broken in the UV once the PQ fields, $\Phi_{1,2} = \rho_{1,2} \,e^{\text{i}a_{1,2}/f_{1,2}} / \sqrt{2} $,
develop a vacuum expectation value (VEV).
We consider the case in which both symmetry breakings happen after inflation. (For more general scenarios, see e.g.~\cite{Lee:2024toz}.)

We also assume that one of the $U(1)$ symmetries, namely $U(1)_2$, is anomalous both under QCD and a dark gauge group which confines at a scale $\Lambda_D$.
Correspondingly, the interactions of the two axions in the broken phase of the SM read\,\footnote{Most of the aspects of our analysis hold if $U(1)_1$ is also charged under the dark gauge group.}:
\begin{align}
\mathcal{L}_a & \supset \frac{\alpha_s}{8 \pi} \left( j_1 \frac{a_1}{f_1} +j_2 \frac{a_2}{f_2} \right) G \widetilde G + \frac{\alpha_{\rm em}}{4 \pi} \left( k_1 \frac{a_1}{f_1} +k_2 \frac{a_2}{f_2} \right)  F \widetilde F 
+ \frac{\alpha_D}{8 \pi} \ell_2 \frac{a_2}{f_2}  G_D \widetilde G_D\,,
\label{eq:Lint}
\end{align}
where $G$ and $F$ are the SM gluon and photon fields, respectively, $G_D$ is the gauge field of the dark sector, $ \alpha_{s} $ is the dimensionless strong coupling constant and $ \alpha_{D} $ is the analogous coupling of the dark sector.
We note that this Lagrangian could naturally arise in the string axiverse context, assuming the kinetic mixing is small.
In this case, an instanton charge can naturally be zero if a $p$-cycle (identified with the axion) does not intersect the brane where the gauge fields propagate.
While the cosmology of string theory axions can be completely different from that of field theory axions, in some scenarios topological defects can still form with similar properties depending on the warping of the extra dimensions~\cite{Benabou:2023npn}.

In this field basis, the anomaly coefficients $j_{1},j_2,\ell_{2}\in \mathbb Z$, while 
$k_i$ are restricted if the global structure of the SM gauge group involves a non-trivial discrete group.
Assuming the quotient factor in Eq.~\eqref{eq:minG} does not act on the dark gauge group, it follows that \cite{Reece:2023iqn,Choi:2023pdp,Cordova:2023her}
\begin{equation}
    \frac{2}{3} j_i + k_i \in \mathbb{Z} ~~~(i=1,2)\,.
    \label{eq:quantization}
\end{equation}

The quotient $\mathbb Z_p$ identifies certain combinations of center elements across $\widetilde{G}_{\rm SM} \equiv SU(3)\times SU(2)\times U(1)$ that act trivially on all SM matter fields when implemented together.  
Such identification depends on $p$.
For example, for $p=3$, one identifies the center of $SU(3)$ with the $\mathbb Z_3$ subgroup of $U(1)_Y$, which constrains the hypercharge.
The minimal choice ($p=6$) corresponds to identifying simultaneously the centers of $SU(3)$ and $SU(2)$ with the $\mathbb Z_6$ subgroup of $U(1)_Y$. 
Such identification allows for gauge transformations which are trivial in $\widetilde{G}_{\rm SM}$, but not in ${G}_{\rm SM}$, defined Eq.~\eqref{eq:minG}, {with $ p \neq 1 $}.
Requiring the action to be invariant under such transformations, after performing an axion shift, forces the quantized conditions in Eq.~\eqref{eq:quantization}.
Since each axion shift can be applied independently, the result for each of the axions in the model must be the same as in the single axion scenario.\,\footnote{We note that these theoretical constraints can be weakened in scenarios with topological mixing, if the discrete group $\mathbb Z_p$ is embedded in both the SM and the dark gauge groups~\cite{Dierigl:2024cxm}.}

In the IR, after both strong sectors confine, the axions develop the following potential:
\begin{align}
V(a_1,\,a_2) = &   \Lambda_{\mathrm{QCD}}^4 \bigg[1- \cos\left(j_1 \frac{a_1}{f_1}  + j_2 \frac{a_2}{f_2}\right)\bigg]
+ \Lambda_D^4 \bigg[1-\cos\left(l_2\frac{a_2}{f_2}\right)\bigg]\,.
\label{eq:V12}
\end{align}
However, throughout the cosmological history, the two axions are expected to develop masses at different times.

If the dark gauge group confines before QCD, $U(1)_2$ is explicitly broken; after integrating the heavy axion $a_2$,  
the flavor basis in Eq.~\eqref{eq:V12} is automatically aligned with the flat direction (associated with the unbroken $U(1)_1$).
If instead QCD confines first, it is easier to identify the flat direction in a different basis, where only one combination of fields couples to gluons:
\begin{align}
   V(a_{G\widetilde G},\,a_{\perp}) = &  \Lambda_{\mathrm{QCD}}^4 \bigg[1- \cos\left(\frac{a_{G\widetilde G}}{f_G}\right)\bigg] 
+ \Lambda_D^4 \bigg[1-\cos\left(\frac{a_{G\widetilde G}}{f_G^\prime}+ \frac{a_{\perp}}{f_\perp}\right)\bigg]\,,
\label{eq:VaGG}
\end{align}
with
\begin{align}
        f_{G}  = \frac{f_1 f_2}{\sqrt{j_1^2 f_2^2 + j_2^2 f_1^2}} \,, \quad
    f_{G}^\prime  = \frac{f_2 \sqrt{j_1^2 f_2^2 + j_2^2 f_1^2}}{f_1 j_2 \ell_2}\,, \quad
    f_\perp  = \frac{\sqrt{j_1^2 f_2^2 + j_2^2 f_1^2}}{j_1\ell_2}\,.
\end{align}
Taking the limit $\Lambda_{\rm QCD} \gg \Lambda_D$, $a_{G\widetilde G}$ can be integrated out, which allows us to identify $f_\perp$ with the effective decay constant of the flat direction; see also~\cite{Choi:2014rja,Choi:2015fiu,Higaki:2016jjh}.

\begin{figure}
    \centering
    \includegraphics[width=0.9\linewidth]{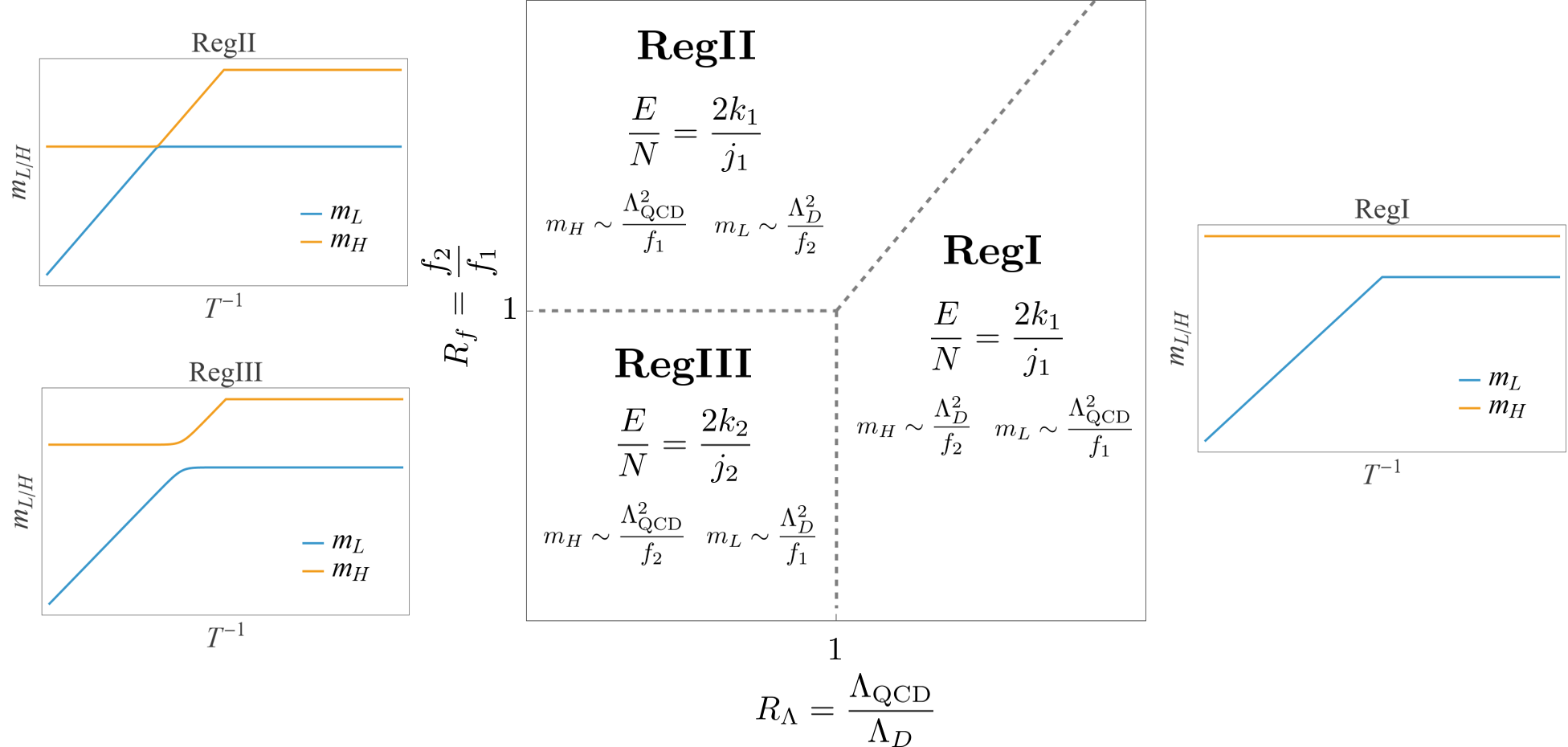}
    \caption{Regimes of parameter space relevant for this work. In the central plot, the dashed lines correspond to $R_f=1$, $R_\Lambda = 1$ and $R_f = R_{\Lambda}^2$; close to these limits, QCD axion coupling quantization is broken. The sub-plots represent the evolution of the eigenvalues with temperature.}
    \label{fig:regimes}
\end{figure}

This discussion suggests the division of the parameter space into three distinct regions to identify the physical properties of the two axions, as indicated in Fig.~\ref{fig:regimes}. These regions are distinguished by different values of the ratio parameters $R_\Lambda \equiv \Lambda_D/\Lambda_{\rm QCD}$ and $R_f \equiv f_2/f_1$, namely:
\begin{equation}
    \begin{aligned}
    \text{RegI:} & \quad R_\Lambda >1~~ \text{and} ~~ R_f < R_\Lambda^2\,;\\
    \text{RegII:} & \quad R_f >1 ~~ \text{and} ~~ R_f > R_\Lambda^2\,;  \\
    \text{RegIII:} & \quad R_\Lambda < 1 ~~ \text{and} ~~ R_f < 1\,.
\end{aligned}
\end{equation}
In each of these regions, the eigenstate identified with the QCD axion satisfies
\begin{equation}
   g_{H,L}\equiv \frac{m_{H,L}^2 f_{H,L}^2}{\Lambda_{\rm QCD}^4}\to 1\,,
\end{equation}
while the other one decouples from the solution to the strong CP problem.\,\footnote{In contrast, for $R_\Lambda \sim 1$, both axions can contribute sizably to the solution to the strong CP problem by sharing similar contributions to the QCD axion sum rule~\cite{Gavela:2023tzu}} We will refer to this field as~\textit{dark axion} and show that it can still impact other aspects of the QCD axion phenomenology.
The coupling of each eigenstate to gluons, $f_{H,L}$, is defined as
\begin{equation}
    \frac{1}{f_{H,L}} \equiv \frac{\left<a_{G\widetilde G}|a_{H,L}\right>}{f_{G}}\,.
\end{equation}

In turn, the physical couplings of each axion to the visible sector read
 \begin{equation}
   L\supset \frac{{a_{H,L}} }{8\pi}
   \frac{\alpha_s}{f_{H,L}} G\widetilde G + \frac{g_{a_{H,L}\gamma\gamma}}{4}  {a_{H,L}} F\widetilde F\,,
\end{equation}
where
\begin{equation}
    g_{a_{H,L}\gamma}= \frac{\alpha_{\rm em}}{2\pi f_{H,L}} \left(C_{H,L} - 1.92\right)\,
    \label{eq:gHL}
\end{equation}
and $a_{H(L)}$ denotes the heavy (light) axion field. 
In Eq.~\eqref{eq:gHL}, $C_{H,L}$ are analogous to the $E/N$ factor but in general they are not quantized for both fields, while the ``1.92" factor is the standard IR piece induced by the mixing of each axion with the neutral pion. 
It reads the same as in the single QCD axion model once the different normalization ($f_{H,L}\neq f_{1,2}$) is taken into account. 
This effect can be computed using the basis adopted in Eq.~\eqref{eq:VaGG}.
In this interaction basis, we can simply perform the standard axion-dependent rotation on quark fields to find the induced coupling of $a_{G\widetilde G}$ to photons, and then project to the mass basis~\cite{Gavela:2023tzu}. 

We now develop the expressions for these couplings. 
In RegI, we find:
\begin{align}
(g_H\,,~g_L) & \approx \left(\mathcal{O}(R_\Lambda^{4})\,,~1\right)\,;\\
    (f_H\,,~f_L) & \approx \left(\frac{f_1}{j_2} R_f\,,~\frac{f_1}{j_1}\right)\,;\\
       \left({\frac{C_H}{\sqrt{g_H}}}\,,~\frac{C_L}{\sqrt{g_L}}\right) & \approx \left(\frac{2 k_2}{\ell_2 R_\Lambda^2}\,,~ \frac{2 k_1}{j_1}\right) \,;\\
       \left(m_H^2\,,~m_L^2\right)& \approx \frac{\Lambda^4_{\rm QCD}}{f_1^2}\left(\frac{\ell_2^2}{R_f^2}  R_\Lambda^4\,,~ {j_1^2}\right)\,.
\end{align}
Instead, in RegII, we have:
\begin{align} 
(g_H\,,~g_L) & \approx \left(1\,,~ \mathcal{O}(R_\Lambda^{-4} R_f^4)\right)\,;\\
    (f_H\,,~f_L) & \approx \left(\frac{f_1}{j_1} \,,~-\frac{f_1 j_1^2}{j_2 \ell_2^2} R_f^3 R_{\Lambda}^{-4}\right)\,;\\
        \left(\frac{C_H}{\sqrt{g_H}}\,,~\frac{C_L}{\sqrt{g_L}}\right) & \approx \left(\frac{2 k_1}{j_1}\,,~\frac{2 j_2}{\ell_2 R_\Lambda^2} \left[\frac{k_1}{j_1} - \frac{k_2}{j_2}\right] \right) \,;\\
        \label{eq:CregII}
       \left(m_H^2\,,~m_L^2\right)& \approx \frac{\Lambda^4_{\rm QCD}}{f_1^2}\left(j_1^2\,,~ \ell_2^2 R_f^{-2} R_{\Lambda}^4\right)\,.
\end{align}
Finally, in RegIII:
\begin{align} 
(g_H\,,~g_L) & \approx \left(1\,,~ \mathcal{O}(R_\Lambda^{-4} )\right)\,;\\
    (f_H\,,~f_L) & \approx \left(\frac{f_1}{j_2}R_f \,,~\frac{f_1 j_2^2}{j_1 \ell_2^2} R_{\Lambda}^{-4}\right)\,;\\
      \left(\frac{C_H}{\sqrt{g_H}}\,,~\frac{C_L}{\sqrt{g_L}}\right)  & \approx \left(\frac{2 k_2}{j_2}\,,~\frac{2 j_2}{\ell_2 R_\Lambda^2} \left[\frac{k_1}{j_1} - \frac{k_2}{j_2}\right] \right)  \,;\\
       \label{eq:CregIII}
       \left(m_H^2\,,~m_L^2\right)& \approx \frac{\Lambda^4_{\rm QCD}}{f_1^2}\left(j_2^2 R_f^{-2}\,,~ \frac{j_1^2 \ell_2^2}{j_2^2}  R_{\Lambda}^4\right)\,.
\end{align}
From these expressions, one infers that 
the dark axion coupling is not quantized in any of the regimes.\,\footnote{Note that the expression $C_i/\sqrt{g_i}$ is equivalent to the photon coupling divided by the mass of each eigenstate.} This agrees with the conclusions in Ref.~\cite{Fraser:2019ojt}. 

In contrast, the QCD axion photon coupling remains quantized within the three regions of parameter space. This does not hold at their boundaries, shown as red dashed lines in Fig.~\ref{fig:regimes}. In particular, for $f_1\sim f_2$ and $R_{\Lambda}\ll 1$ with arbitrary anomaly coefficients, quantization is lost. 
Lighter axion-like fields can thus significantly affect the quantization of the QCD axion coupling, as the field combination coupling to gluons generically differs from that coupling to photons, unless symmetry constraints enforce otherwise.
For example, if the SM is embedded in a GUT group that commutes with $U(1)_{1,2}$, the QCD axion coupling remains quantized even in this $R_f\to 1$ limit. 
Moreover, under this assumption, the leading term to $C_L/\sqrt{g_L}$ vanishes for the dark axion in RegII and RegIII, {placing it} far from the QCD axion photon band.

Finally, we note that even though the QCD axion coupling is quantized, the $E/N$ factor reads differently across the three regions identified in Fig.~\ref{fig:regimes}. 
To constrain its minimal value, we should first discuss what are the requirements on the anomaly coefficients in order to solve the DW problem in our model.

\section{Domain wall number constraint}\label{sec:DWnumber}

After the spontaneous breaking of one of the global $U(1)$ symmetries, each causal patch of the universe picks a PQ VEV with a random phase.
The non-vanishing winding number of the axion field along a closed loop $ C $ between these patches, $ \frac{1}{2\pi} \oint_{C} d\theta \neq 0$, implies the existence of a singular point where the PQ symmetry is restored. 
This signals the presence of cosmic strings, according to the Kibble-Zurek mechanism~\cite{Kibble:1976sj,Zurek:1985qw}.

Once an axion mass is generated, one of the original global symmetries is broken to a discrete group giving rise to multiple degenerate vacua and leading to the formation of DWs.
If these DWs are long-lived, the energy density of the network dilutes much more slowly than radiation and would come to dominate the energy density of the universe,\,\footnote{See Ref.~\cite{Hong:2025piv} for the gravitational wave spectrum in the presence of temporal DW domination.} posing a cosmological problem.
If DWs are formed after inflation, one of the most simplest solutions is to require $N_{\rm DW} = 1 $,\,\footnote{There are other ways to make the DWs decay, but solutions beyond $ N_{\rm DW} =1$ typically lead to other difficulties~\cite{Lu:2023ayc}.
For instance, adding a bias term in the QCD axion potential generically induces the axion quality problem~\cite{Caputo:2019wsd,Cheong:2022ikv}.
DWs can also disappear due to nucleation into strings~\cite{Vilenkin:1982ks,Kibble:1982dd}.
However, this is highly suppressed unless the scales involved are fine tuned.}
in which case all DWs are bounded by strings and rapidly annihilate.
We now discuss the conditions under which this is achieved.

The solutions $\theta_i\equiv a_i/f_i$ that minimize the potential are:
\begin{equation}
    \begin{pmatrix}
        \theta_1 \\ \theta_2 
    \end{pmatrix} = 2 \pi \mathcal{A}^{-1} \begin{pmatrix}
        m_1 \\m_2 
    \end{pmatrix}\,,\quad \text{with} \quad  \vec m \in \mathbb{Z}^2\,,
\end{equation}
and
\begin{equation}
    \mathcal A = \begin{pmatrix}
        j_1 & j_2 \\
        0 & \ell_2
    \end{pmatrix}\,.
\end{equation}
These equations select a discrete, periodically repeating set of minima in field space.
From the $2\pi$ periodicities of $\theta_{1,2}$, two minima $\vec \theta, \, \vec \theta^\prime$ associated to integer vectors $\vec m, \, \vec m^\prime \in \mathbb Z^2$ are equivalent if they represent the same point in the torus (the field space of two periodic fields), i.e.
\begin{equation}
    \vec \theta - \vec \theta^\prime \in 2 \pi \mathbb Z^2\quad \Leftrightarrow \quad  \vec m -\vec m^\prime \in \mathcal A \mathbb Z^2\,.
\end{equation}
Distinct physical vacua on the torus are therefore in one-to-one correspondence with equivalent classes of integer vectors modulo $\mathcal A \mathbb Z^2$.
It follows that the total number of distinct minima in $T^2$ is given by the dimensionality of $ \vert \mathbb Z^2/\mathcal A \mathbb Z^2 \vert = \vert \text{det}\,\mathcal A \vert$. This holds for more general square matrices and agrees with previous studies, e.g.~\cite{Kondo:2025hdc}.

The same conclusions can be found using the Smith normal form, which counts the number of minima along a $2\pi$ trajectory of the following $ \theta_{p_{1,2}} $ fields:
\begin{align}
   p_1 \theta_{p_1} & = {j_1 \theta_1 + j_2\theta_2} \nonumber \\
   p_2 \theta_{p_2} & = {\ell_2\theta_2}
   \label{eq: periodicity basis}
\end{align}
with ${p_1 = \text{gcd}(j_1,j_2,\ell_2)}$ and $p_2 = |{\rm det} \, \mathcal A|/p_1$. The number of minima is then given by $p_1 \,p_2$.\,\footnote{The determination of the number of minima using this method can be easily generalized to an arbitrary number of axions and instantons. In general, $N_{\rm total}$ is simply given by the multiplication of all the elements of the Smith normal form.}

\begin{figure}[t]
    \centering
    \includegraphics[width=0.4\linewidth]{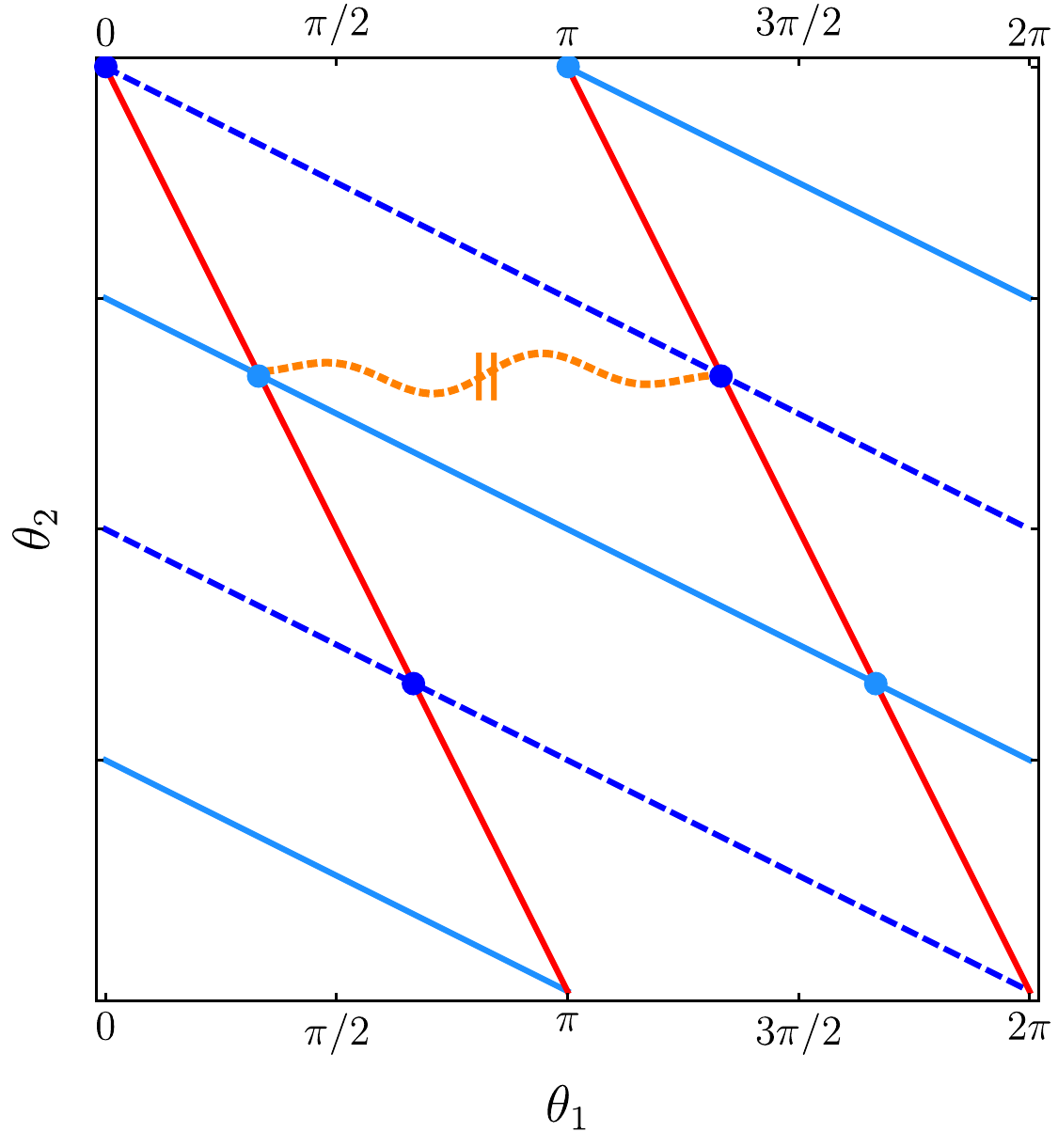}
    \caption{The domain wall number as seen by two periodic flat directions (blue dashed and solid) versus the total domain wall number of a two-axion system. Here, we chose ${\mathcal{A} = \{\{2,\,4\},\,\{2,\,1\}\}}$.}
    \label{fig:minima}
\end{figure}

It is important to distinguish this number, 
${N_{\rm DW} = |{\rm det} \,\mathcal A|}$, from the number of DWs attached per string, ${N_{\rm string} = |{\rm det} \,\mathcal A|/{\rm gcd}(j_1,j_2)}$ \cite{Benabou:2023npn,Lee:2024toz}.
While the former is independent of which gauge group confines first, the latter is not. As clarified in a recent paper~\cite{Kondo:2025hdc}, it is not enough to guarantee $N_{\rm string}=1$ to avoid a cosmological catastrophe.
To see this, we give a concrete example of the minima structure in Fig.~\ref{fig:minima}. 
Assuming that QCD confines first,
the flat directions satisfy the constraint
\begin{align}
    j_1 \theta_1 + j_2 \theta_2 =2\pi m_1\,,
    \label{eq:disconnectedmin}
\end{align}
with 
\begin{equation}
   m_1 =( 0,\,\dots,\,{\rm gcd} (j_1,j_2){-1})\,.
\end{equation}
For $m_1$ larger than the largest value in this set, the difference in winding number of the massless axion around a string is equivalent to zero, due to the periodicity of the theory.
Hence, there are ${\rm gcd} (j_1,j_2)$ topologically distinct windings ($\Delta \theta_L\neq 0$) around strings.

In the example of Fig.~\ref{fig:minima}, ${\rm gcd} (j_1,j_2) =2$; hence there are two distinct winding configurations represented by the solid and dashed blue lines.
In this example, each string configuration goes through three distinct minima, so it has attached three DWs.
Moreover, DWs that do not end on strings can also form, connecting horizons that settle in minima associated with disconnected lines on the torus (such a DW is represented by an orange wiggle in the figure).

If $N_{\rm string} = 1$, the string network quickly collapses once the dark potential turns on, but this does not prevent the formation of DWs associated to disconnected minima lines.
To prevent these structures from causing a cosmological problem, one must require $N_{\rm DW} = 1$. 
This condition ensures there are no stable DWs after the second potential confines.
However, if $N_{\rm string} \neq 1$, the DW network created at $T\sim \Lambda_{\rm QCD}$
would survive until $\Lambda_D$, which can also pose a cosmological problem \cite{Lee:2024toz}. 
This does not occur in our model with $ \ell_1 = 0 $, since
\begin{equation}
N_{\rm DW} = 1 \quad \Leftrightarrow  \quad j_1=\ell_2 =1\,,
\end{equation}
while $j_2 {\in \mathbb{Z}}$ is unconstrained. This condition implies $N_{\rm string} = 1$.
We assume this holds from here on.

\section{Minimum photon coupling}\label{sec:minc}
\begin{table}[t]
    \centering
   \begin{tabular}{|c|c|c|c|} 
   \hline
    $j_{2}$ & $k_{2}$ & $E/N$ & $\vert E/N-1.92 \vert$ \\
    \hline
    1 & 4/3 & 8/3 & 0.75 \\
    2 & 5/3 & 5/3 & 0.25 \\
    3 & 3 & 2 & 0.08 \\
    4 & 13/3 & 13/6 & 0.25 \\
    5 & 14/3 & 28/15 & 0.053 \\
    \hline
\end{tabular}
   \begin{tabular}{|c|c|c|c|} 
   \hline
    $j_{2}$ & $k_{2}$ & $E/N$ & $\vert E/N-1.92 \vert$ \\
    \hline
    6 & 6 & 2 & 0.08 \\
    7 & 19/3 & 38/21 & 0.11 \\
    8 & 23/3 & 23/12 & 0.0033 \\
    9 & 9 & 2 & 0.08 \\
    10 &28/3 & 28/15 & 0.053 \\
    \hline
\end{tabular}
\begin{tabular}{|c|c|c|c|} 
   \hline
    $j_{2}$ & $k_{2}$ & $E/N$ & $\vert E/N-1.92 \vert$ \\
    \hline
    11 &32/3 & 64/33 & 0.019 \\
    12 &12 & 2 & 0.08 \\
    13 &37/3 & 74/39 & 0.023 \\
    14 &41/3 & 41/21 & 0.032 \\
    15 &14 & 28/15 & 0.053 \\
    \hline
\end{tabular}
    \caption{Values of $ E/ N$ that minimize the QCD axion coupling to photons in RegIII.}
    \label{tab:quantization}
\end{table}

We now derive the minimal QCD axion photon coupling given the constraints on the anomaly matrix derived in the previous section.
In RegI and RegII, we find that the prediction for this minimal coupling is the same as in the single axion model: it arises for $E/N = 8/3$. 
However, this does not hold in RegIII since $E/N$ depends on $j_2$ and is therefore not fixed.
Tab.~\ref{tab:quantization} shows the values of $E/N$ that minimize the QCD axion photon coupling for different values of this anomaly coefficient. In all cases, the model remains free of the DW problem.
The results show that there is some periodicity associated to the value of the minimal coupling.
 For instance, for $j_2 = 3n \,\,{\leq 12} \,,~n\in\mathbb Z$, the minimum occurs for $E/N = 2$.
 This is an interesting benchmark for DFSZ-inspired models.
Consider $a_2$ as 
 the QCD axion in a standard DFSZ model: 
$j_2$ is then a multiple of 3, since  the SM quarks are charged under $U(1)_{2}$, which causes a DW problem.
The inclusion of a second axion $a_1$ with $j_1=1$
straightforwardly resolves this issue~\cite{Lee:2025zpn}.
We find that a consequence of this mechanism is that the QCD axion coupling can become further suppressed, even when constrained by the minimal SM gauge group structure.

Altogether, this analysis shows that an axion signal observed to the right of the $E/N =8/3$ line could be perfectly compatible with the minimal SM gauge group setting and a post-inflationary cosmology, safe from a DW problem.
Searches beyond this theoretical benchmark are motivated and, in some cases, necessary to probe multi-axion models.

Even in the parameter space where the minimal model prediction for $E/N$ is recovered, a dark axion can still impact the dark matter fraction of the QCD axion.
As the most sensitive -- haloscope -- experiments searching for the electromagnetic coupling rely on this fraction, it is important to quantify this effect in order to 
re-interpret experimental projections.

\section{Dark matter abundances}\label{sec:dm}

The dark matter abundance of the two axions in the post-post inflationary scenario includes the contributions from the standard misalignment mechanism and axion emission from cosmic strings.

\subsection{Misalignment}

Once an axion starts oscillating with $ { m_{i}(T_{H,L}^{\rm osc})\approx 3H(T_{H,L}^{\rm osc}) }$, its number density is given by
\begin{align}
    n_{H,L}^{\rm WKB}(T_{H,L}^{\rm osc})  \approx \frac{1}{2} m_{H,L}(T_{H,L}^{\rm osc}) \,a_{H,L}^2 (T_{H,L}^{\rm osc})\,,
\end{align}
under the WKB approximation. 
Here, 
the initial field amplitude is computed for $\langle \theta_{1,2}^{2} \rangle \approx \pi^{2}/3$ which corresponds to the {ensemble} average of the axion VEV in causally disconnected regions (up to anharmonic effects, which we neglect).
The temperature dependence of the eigenvalues is implicit in the mass matrix,
\begin{equation}
    M^2 = \begin{pmatrix}
        \dfrac{m_{G\widetilde G}(T)^2 R_f^2}{1+R_f^2} &  \dfrac{m_{G\widetilde G}(T)^2 R_f}{1+R_f^2} \\
         \dfrac{m_{G\widetilde G}(T)^2 R_f}{1+R_f^2} &  \dfrac{m_{G\widetilde G}(T)^2 }{1+R_f^2} + m_{G_D\widetilde G_D}^2 
    \end{pmatrix}\,,
\end{equation}
where
\begin{align}
    \frac{m_{G\widetilde G}(T)^{2}}{m_{{G\widetilde G},0}^{2} }\approx \text{min} \left\{1,\,\left(\frac{T}{T_{\rm QCD}} \right)^{-2 n}\right\}
    \,,
\end{align}
with $T_{\rm QCD} \approx 100$\, MeV,  $n\approx 3.34$\,\footnote{We use the value of this parameter adopted in Ref.~\cite{Dunsky:2025sgz}, even though there are other predictions from distinct lattice studies.} and $m_{G\widetilde G,0}^2 = {\Lambda_{\rm QCD}^4 }/{f_G^2}$. 
We assume, for now, ${m_{G_D\widetilde G_D}^2 = \Lambda_D^4/f_2^2}$ %
is constant.
Departures from this assumption are explored in Sec.~\ref{sec:pheno}.

To relate the number density at $T^{\rm osc}$ with that of the present time, one must consider the possibility of level crossing. 
This can happen in RegII as long as the crossing temperature $T_\times < T_{H,L}^{\rm osc}$.
At the crossing temperature, the difference between the eigenvalues is minimal and the two axion flavors are maximally mixed.
If the timescale of the resonance in mixing angle, $\zeta$, is large in comparison to the oscillation frequency of the two axions, the number densities of the two eigenstates are independently conserved.
This adiabatic regime corresponds to the parameter space where~\cite{Dunsky:2025sgz}
\begin{equation}
    \begin{aligned}
    \gamma \equiv \left| \frac{m_H - m_L}{2\dot{\zeta}}\right|_{T=T_\times}  
    = R_f^{-2/n} \bigg[\frac{m_{{ G\widetilde G},0} \,R_{\Lambda}^{2+4/n}}{H(T_{\rm QCD})\, n \,R_f^3} + \mathcal{O} \left(R_f^{-5}\right)\bigg] > 1\,.
    \end{aligned}
\end{equation}
In such regions, to get the number densities today, one just multiplies by the redshift factor:
\begin{align}
    n_{H,L}^{{\rm WKB},0} = n_{H,L}^{\rm WKB} (T_{H,L}^{\rm osc}) \left( \frac{a_{0}}{a_{\rm osc}} \right)^{-3} \,,
\end{align}
and obtains the energy density as $\rho_{H,L} ^0\approx m_{H,L}^0 n_{H,L}^0$.
If instead, the level crossing is non-adiabatic ($\gamma < 1$), resonant conversation between the eigenstates must be taken into account:
\begin{align}
n_{H,L} (T<T_{\times})  = {(1-P_{\rm LZ})} n_{H,L}^{\rm WKB} + {P_{\rm LZ}} n_{L,H}^{\rm WKB}\,,
\label{eq:LZmod}
\end{align}
with ${P_{\rm LZ}=e^{- \pi \gamma/2}}$ given by the Landau-Zener formalism~\cite{Dunsky:2025sgz}.

\subsection{Cosmic Strings}

\begin{figure}[t!]
    \centering
    (a)
    
\includegraphics[width=0.9\linewidth]{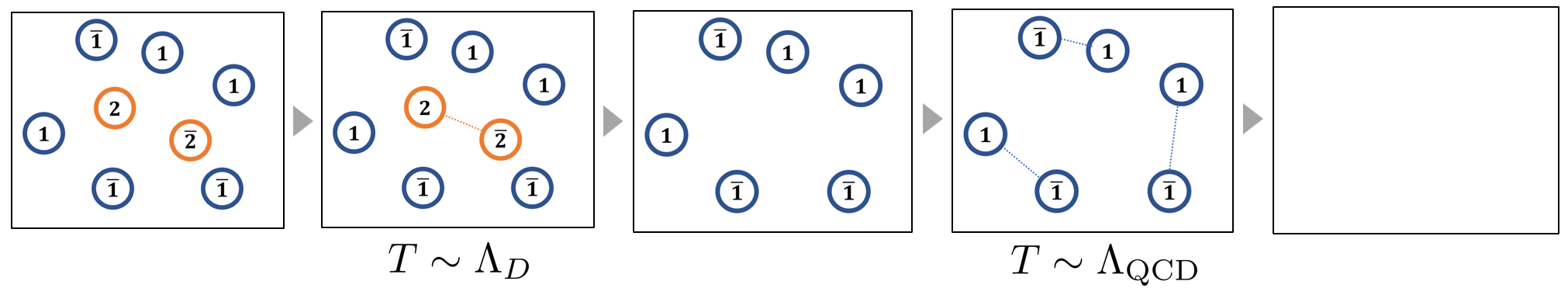}
    
    (b)
    
\includegraphics[width=0.9\linewidth]{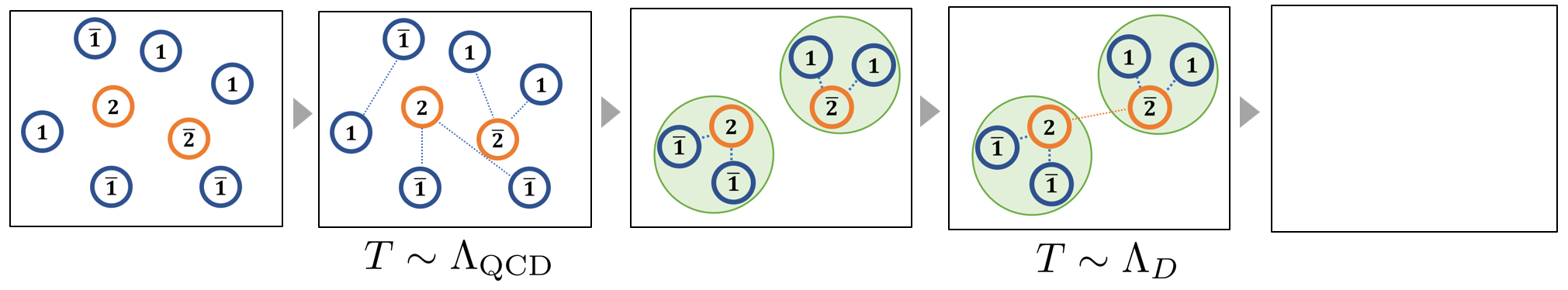}
    \caption{Schematic illustration of cosmic string network evolution when (a) the dark potential dominates first, or (b) the QCD potential dominates first. In the latter case, cosmic bundle structures emerge.
    Here, we assume $j_2 = 2$.}
    \label{fig:cosmic_string}
\end{figure}
\begin{figure}[t]
  \centering
  \includegraphics[width=0.48\textwidth]{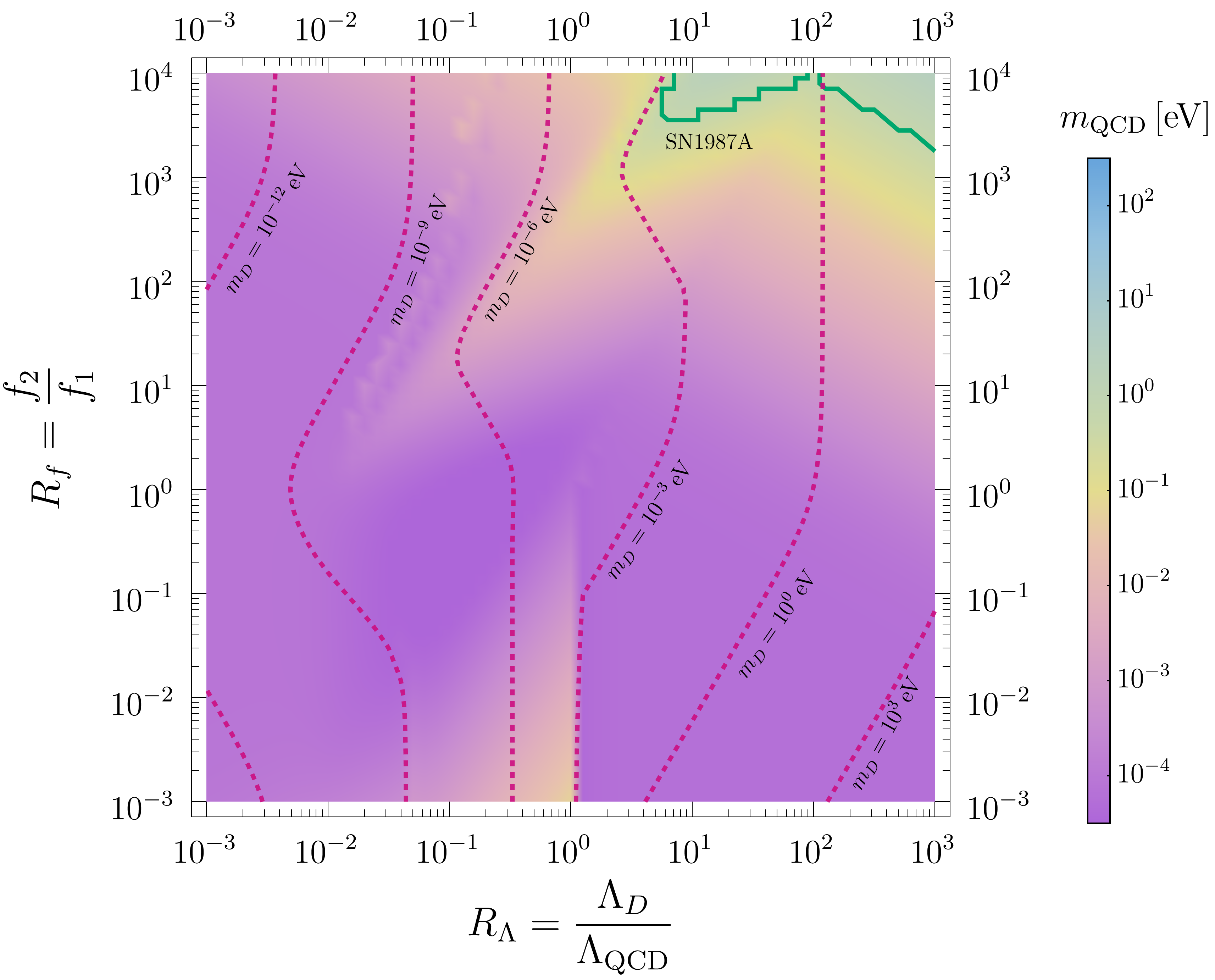}\hfill
  \includegraphics[width=0.48\textwidth]{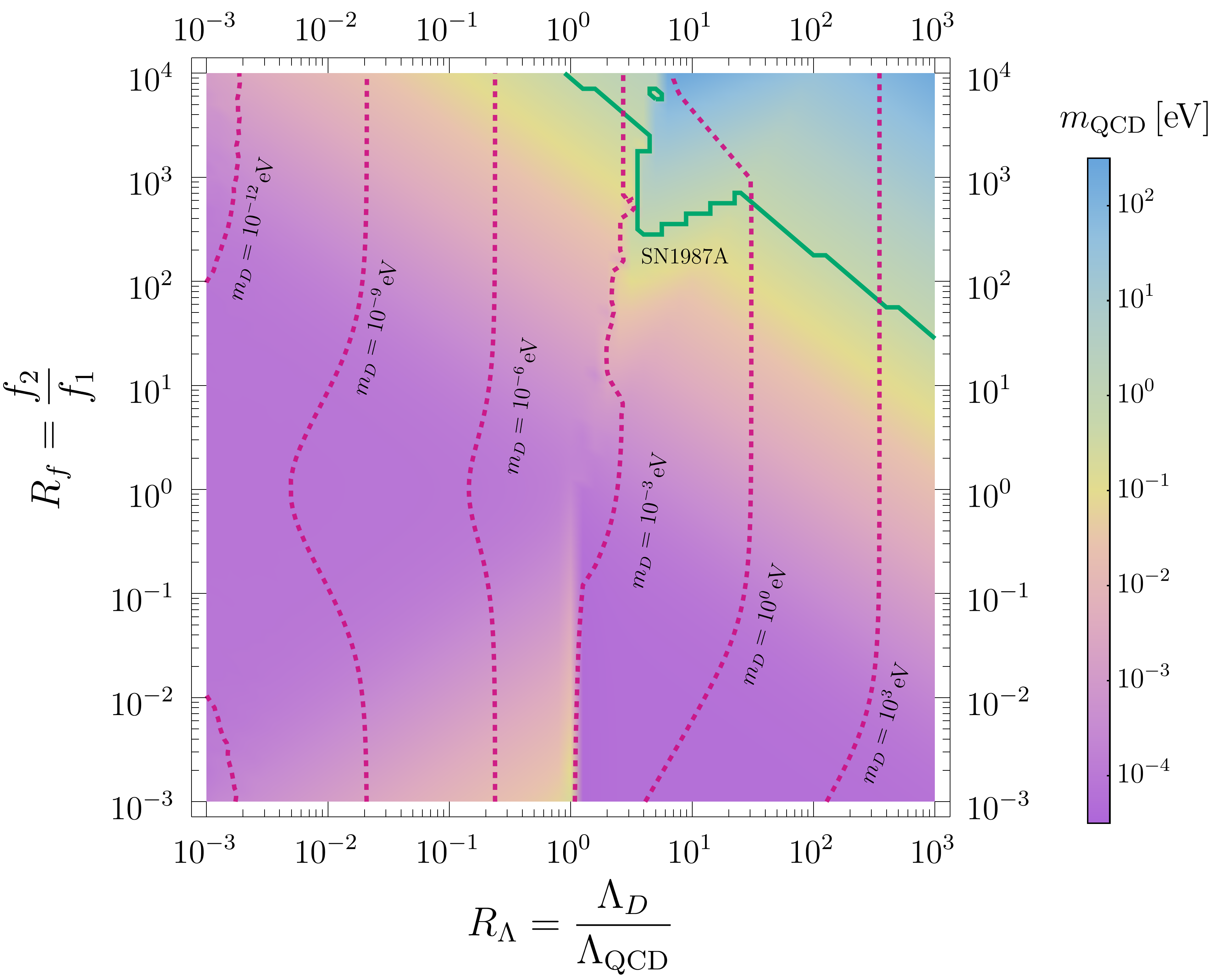}
    \includegraphics[width=0.48\textwidth]{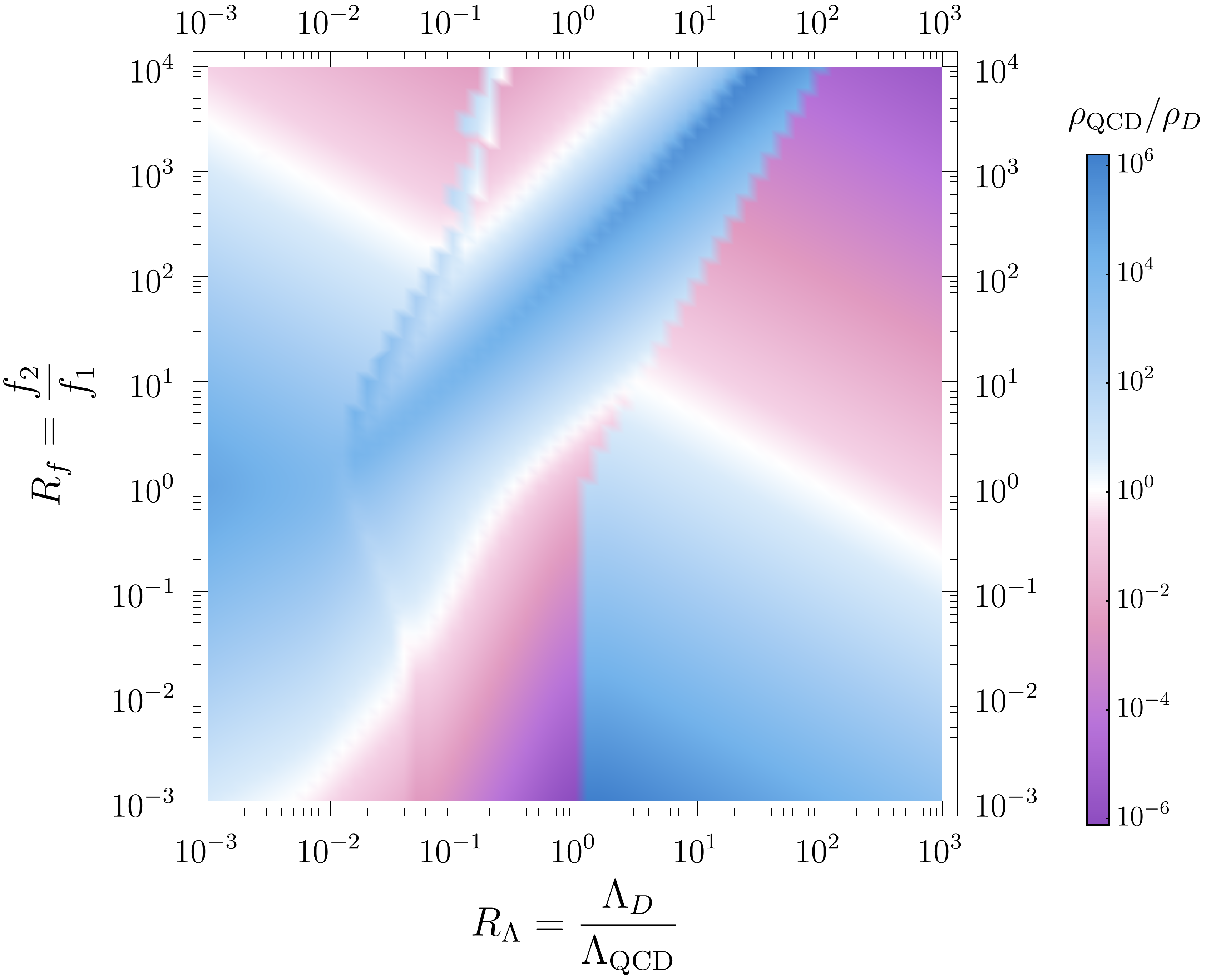}\hfill
  \includegraphics[width=0.48\textwidth]{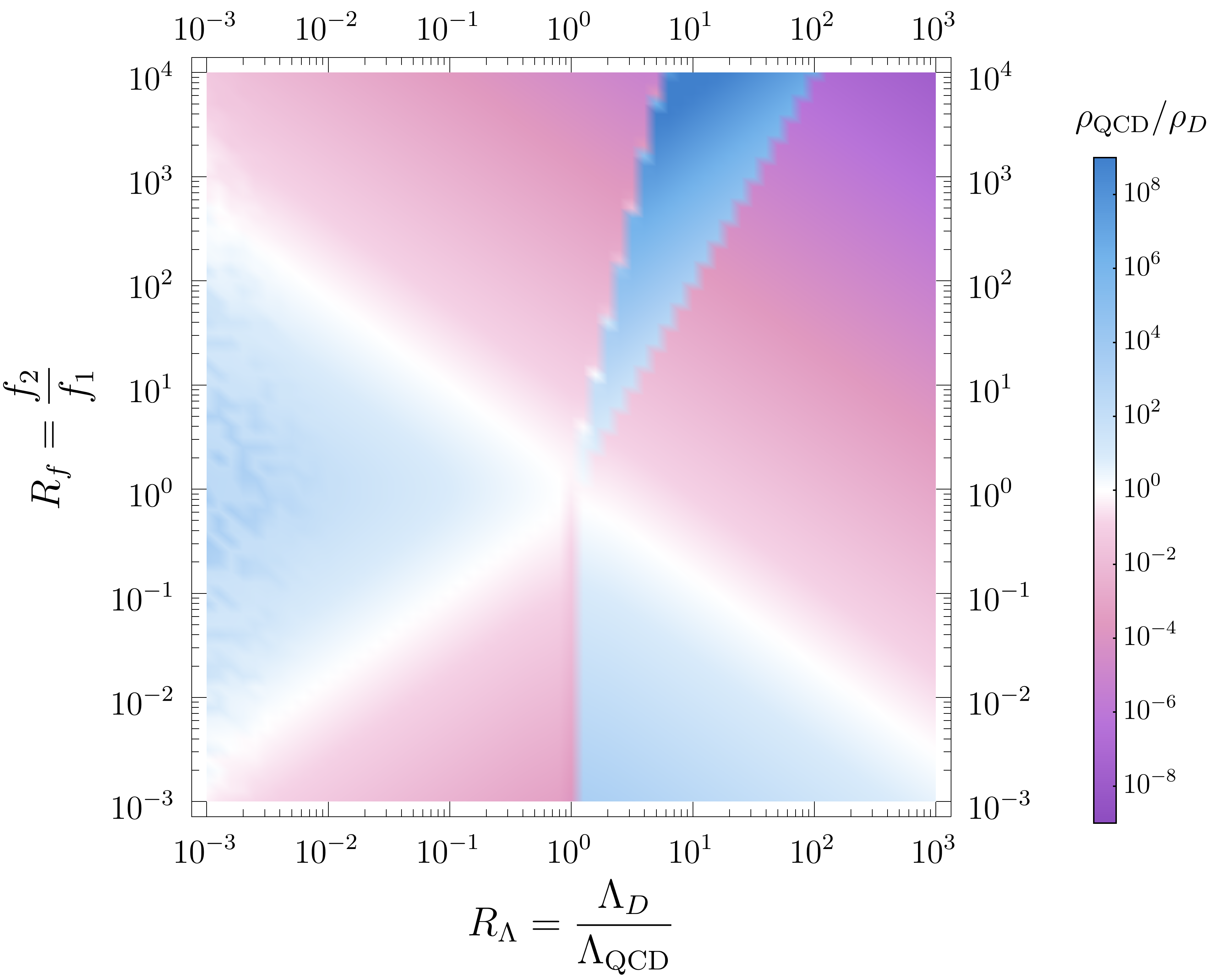}
  \caption{The parameter space in which the two axions explain the totality of dark matter observed, for $j_2=1$. The figures on the top (bottom) show the variation of the axion masses (ratio of dark matter abundances) across this parameter space. %
  The region enclosed by the green line is excluded by supernovae bounds~\cite{AxionLimits}.  In the left (right) panels, the dark axion mass is assumed temperature-independent, while in the right panels it follows a power law with the same exponent as the QCD axion mass.}
    \label{fig:masspredictions}
\end{figure}

In the UV, there are two species of cosmic strings associated with the global shift-symmetries of $a_{1,2}$ (hereafter referred to as ${1,2}$-strings).
Assuming the dark potential confines first, at $T\sim \Lambda_D$, $U(1)_2$ is broken and DWs connecting 2-strings begin to form and rapidly annihilate, while $U(1)_1$ remains exact.
When $T\sim \Lambda_{\rm QCD}$, DWs also form between 1-strings and the network collapses, given our constraints on the anomaly coefficients.
We make a schematic illustration of the network evolution in Fig.~\ref{fig:cosmic_string} (a).

In this two-axion setup, the cosmic string dynamics decouple into two independent single-axion networks.
Each network comprises strings with tension
\begin{equation}
\mu_{1,2} \simeq \pi f_{1,2}^{2} \log(m^r_{1,2}/H) \,,
\end{equation}
where $ m^r_{1,2} \sim \sqrt {\lambda} f_{1,2}$ is the mass of the corresponding PQ radial mode (we take {the quartic coupling of the PQ scalar field} $\lambda\sim 1$). 

Within each network, the string density evolves to reach the scaling solution, where the number of strings per Hubble patch is approximately $\xi \sim \mathcal O(1)$. (Recent simulations found that this parameter grows logarithmically at most~\cite{Gorghetto:2018myk,Gorghetto:2020qws,Buschmann:2021sdq,Saikawa:2024bta,Kim:2024wku}, even though there is no clear consensus~\cite{Correia:2024cpk}).\,\footnote{We  also note there are $O(1)$ uncertainties in the computation of the axion yield from the annihilation of the string-domain wall network~\cite{Hiramatsu:2012gg,Kawasaki:2014sqa,Benabou:2024msj}, which we neglect in this work.}
The network energy density is then given by
$ \rho_{1,2} \approx \xi \,\mu_{1,2} H^2 $, which dissipates mainly through axion emission.

While the prediction for the total energy density of the radiated axions is {relatively} robust, there is not an exact agreement regarding the energy spectrum.
Depending on the simulation prescription, the spectrum can be either scale-invariant ($q=1$) \cite{Buschmann:2021sdq} or IR dominated ($q>1$) \cite{Gorghetto:2018myk,Gorghetto:2020qws,Kim:2024wku}, with most of the energy being emitted with momenta of
order Hubble.
This leads to the following predictions for the number density of axions produced from string emission, in comparison to the misalignment contribution:
\begin{align}
    \frac{n_i^{\rm (str.)}}{n_{i}^{\rm (mis.)}} & \approx \frac{16 H (T_i^{\rm osc}) \xi \mu_i}{\left<\theta_i^2 \right> m_i(T_i^{\rm osc}) f_{i}^2} \begin{dcases}
        \log^{-1} [m^{r}_{i}/H(T_i^{\rm osc})]  & {(q=1)} \\
        1 & {(q>1)} 
    \end{dcases} \nonumber \\
    & \approx \frac{16 \xi}{\pi} \begin{dcases}
        1 & {(q=1)} \\
        \log [m^{r}_{i}/H(T_i^{\rm osc})] & {(q>1)} 
    \end{dcases}\,.
    \label{eq:spectrum}
\end{align}
We choose $q=1$ for our estimates.
Since the IR-dominated spectrum enhances this prediction, the QCD axion mass which explains dark matter is slightly larger in this case.

We now turn to the scenario where the QCD potential dominates first (${R_\Lambda \ll 1}$).
In this case, the analysis is more intricate as the PQ basis is not aligned with the eigenbasis, which results in the formation of \emph{string bundles} \cite{Higaki:2016jjh,Eto:2023aqr,Lee:2024toz}.
Once QCD confines, a linear combination of $U(1)_1$ and $U(1)_2$ is broken, corresponding to the shift-symmetry of $a_{G\widetilde G}$ in the basis of Eq.~\eqref{eq:VaGG}.
Therefore, when $T\sim T_{\rm QCD}$, two types of DWs form: those connecting 1- or 2-strings to their own anti-strings, and those connecting 1-strings to 2-strings.
The string configurations around which $\Delta \theta_H\neq  0$  quickly annihilate since the DW number associated to the largest potential in our model is 1; see Sec.~\ref{sec:DWnumber}.
The tension associated to these string configurations is set by the largest PQ mass inducing the dominant string tension.
If the QCD axion is mostly identified with $a_1$, then $\mu_H \propto f_1^2 \sim f_G^2$. 

In turn, the string configurations associated to $\Delta \theta_H = 0$ ($\Delta \theta_L\neq  0$) survive until the dark potential turns on. These string bundles consist of one 1-string connected to $j_2$ anti 2-strings; see Fig.~\ref{fig:cosmic_string} (b).
At $T\sim T_D$, DWs also form between these string bundles, which causes the string network to collapse.
In the scenario where $a_{G\widetilde G}\approx a_1$, the light field is $a_L\approx a_2$ so $f_2$ sets the tension of the string bundle, which in general is given by $\mu_L \sim f_\perp^2$.

We clarify that the string tensions in this model are not exactly $f_G^2$ or $f_\perp^2$.
This would hold if $a_{G\widetilde G}$ and $a_{\perp}$ were the angular modes associated to the PQ fields in the UV, in which case there would be no confining force from the DWs attached to strings. 
The differences between such setup and our UV model are encoded in the composite nature of the string bundle, which should be relevant close to the string core, though we expect the associated corrections to be small.\,\footnote{This might not hold in a model with more axions coupled to gluons.
Let us assume $f_3\ll f_{1,2}$. In this case, $f_G \sim f_3$ which would not characterize in general the tension of string configurations made out of 1- and 2-strings, which could still form in this context.
We leave for future work a systematic study of the dynamics of string bundles in $N>2$ axion scenarios.}
Therefore, the overall systematic error in our results follows that of the single post-inflationary axion model.

\subsection{Combined predictions}

Given these considerations, we compute the relic abundance of the QCD axion by requiring that both axions explain the totality of dark matter observed in the universe.
The results are shown in the left panels of Fig.~\ref{fig:masspredictions}, assuming $j_2 = 1$. These can be easily interpreted for different values of this parameter, by reading the $y$-axis as $R_f/j_2$. 
A larger $j_2$ would further reduce the parameter space in RegIII where the QCD axion is the dominant dark matter field.

We observe that RegI and RegIII feature extensive parameter regions where the QCD axion becomes subdominant. 
In the complementary regions, the model reduces to the single QCD axion scenario: the dark axion either becomes massless or has a suppressed decay constant that makes its contribution to the relic abundance negligible.

In RegII, due to level crossing, two interesting scenarios occur.
When this crossing is adiabatic, the QCD axion can explain the totality of dark matter with a different decay constant, hence a different mass value. In this regime, such phenomena makes the QCD axion more visible to experiments in comparison to the single QCD axion model. In fact, supernovae bounds already exclude some of this parameter space.
However, when the crossing is non-adiabatic, efficient resonant conversation takes place\,\footnote{We include the LZ conversion probability in the number density of axions emitted by the string network.
Even though the derivation of this probability assumes spatially homogeneous axion fields~\cite{Dunsky:2025sgz}, it remains valid as long as the momenta of the axions emitted by the string are small in comparison to the around the crossing time. This is true due to the redshift effect, unless $T_{\times} \approx T^{\rm osc}$.} which can make the QCD axion under-abundant.
This occurs in the parameter space where $R_f\gg 1$ and $R_\Lambda \lesssim 1$ in Fig.~\ref{fig:masspredictions}.
The large variation of masses around $R_\Lambda\sim 10^{-1}$ is due to large gradients in the field amplitude when $T^{\rm osc}\sim T_{\times} $.

An important assumption of our framework is that the dark axion mass is approximately constant across the parameter space.
More generally, one expects the dark axion to acquire a temperature-dependent mass of the form $m_{G_D\widetilde G_D}\propto (T/\Lambda_D)^{-\alpha}$, as occurs for a dark gauge group $SU(N)_D$ with a similar number of flavors as QCD that shares a common temperature with the SM bath.
To compare this with our previous scenario, we take $\alpha\approx n_{\rm QCD}$ for illustration and re-compute the dark matter abundances. 
The results are shown in the right panels of Fig.~\ref{fig:masspredictions}.

An immediate consequence of this new setup is that level crossing in RegII is now restricted to the region where $R_\Lambda > 1$.
Moreover, mixing not only causes the QCD axion to misalign with the dark axion decay constant, but also delays the onset of its oscillations. As a result, the QCD axion is less diluted by redshift compared to the previous scenario, requiring a larger mass to account for the observed relic abundance. 
Supernovae constraints already exclude large portions of this region.
In other regions of parameter space, a temperature-dependent dark axion mass instead renders the dark axion the dominant component through a delayed onset of its oscillations.

We note that including a logarithmic scaling of $\xi$ would enhance the number density of the QCD axion, requiring a smaller decay constant to explain the dark matter abundance. Consequently, the QCD axion would become more visible, with a mass roughly one order of magnitude larger, while the regions where it dominates remain practically unchanged. Our results are therefore conservative in this respect.

\section{Phenomenological constraints}\label{sec:pheno}
\begin{figure}[t]
  \centering
  \includegraphics[width=0.48\textwidth]{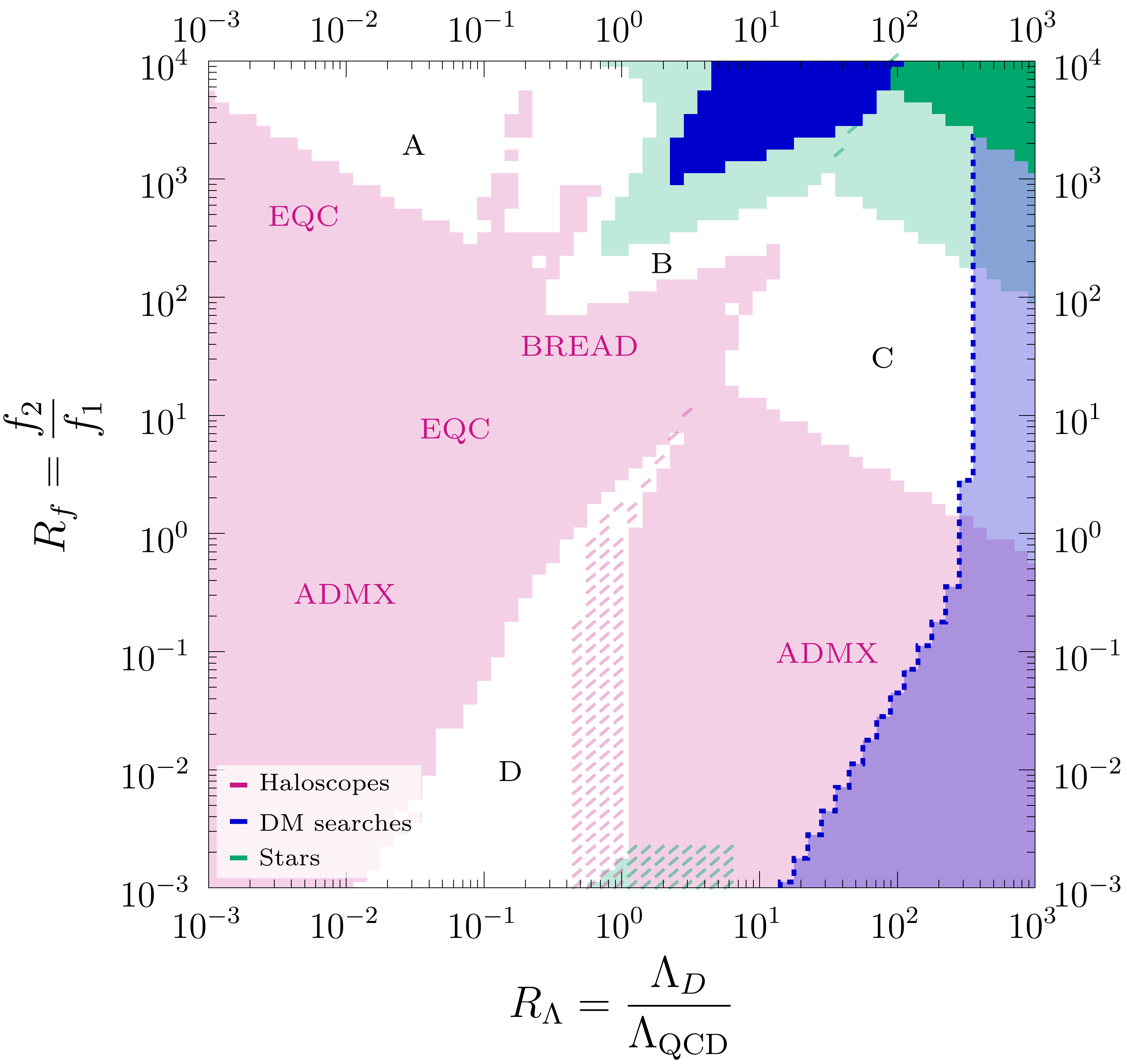}\hfill
  \includegraphics[width=0.48
  \textwidth]{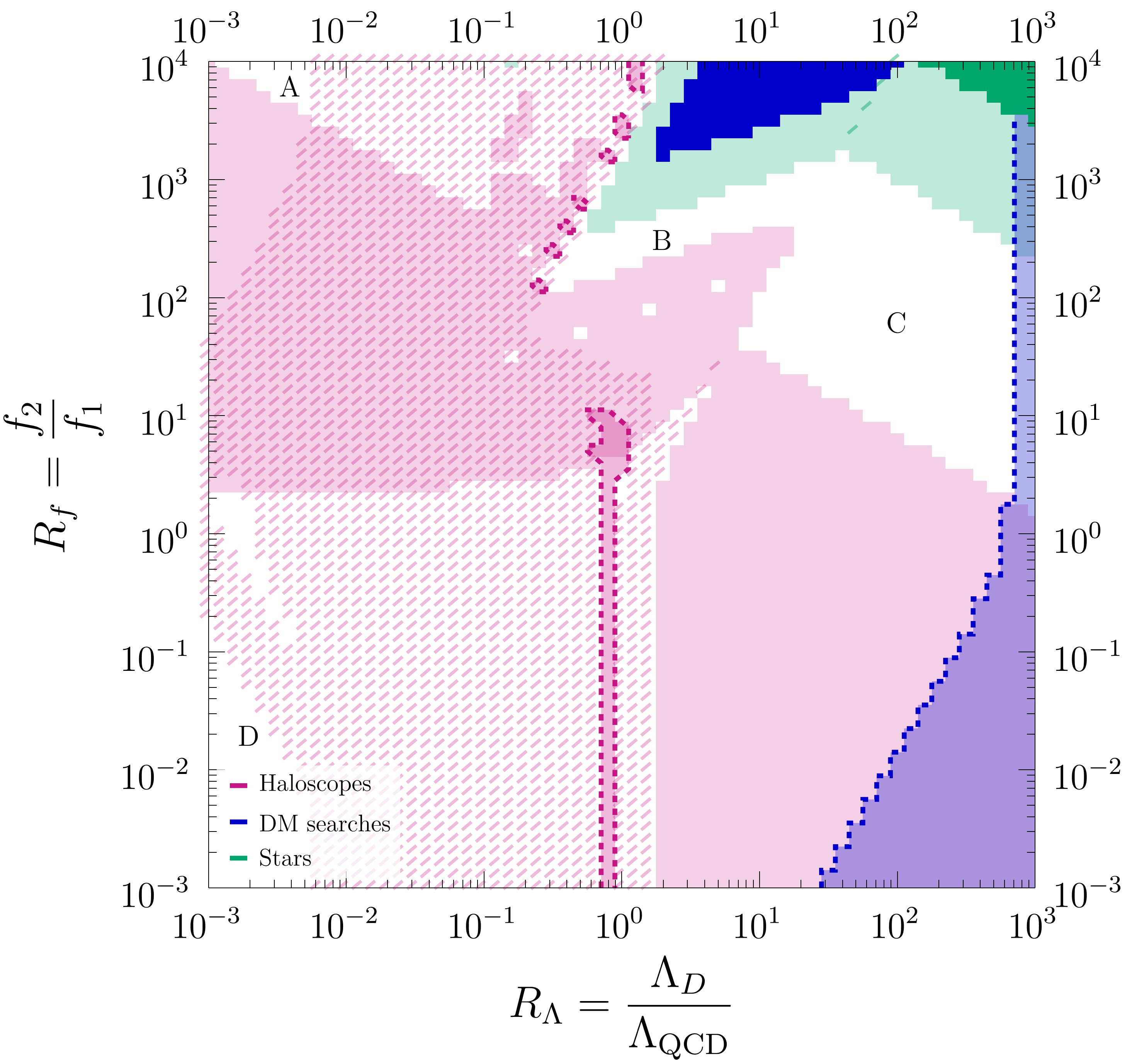}
  \includegraphics[width=0.48\textwidth]{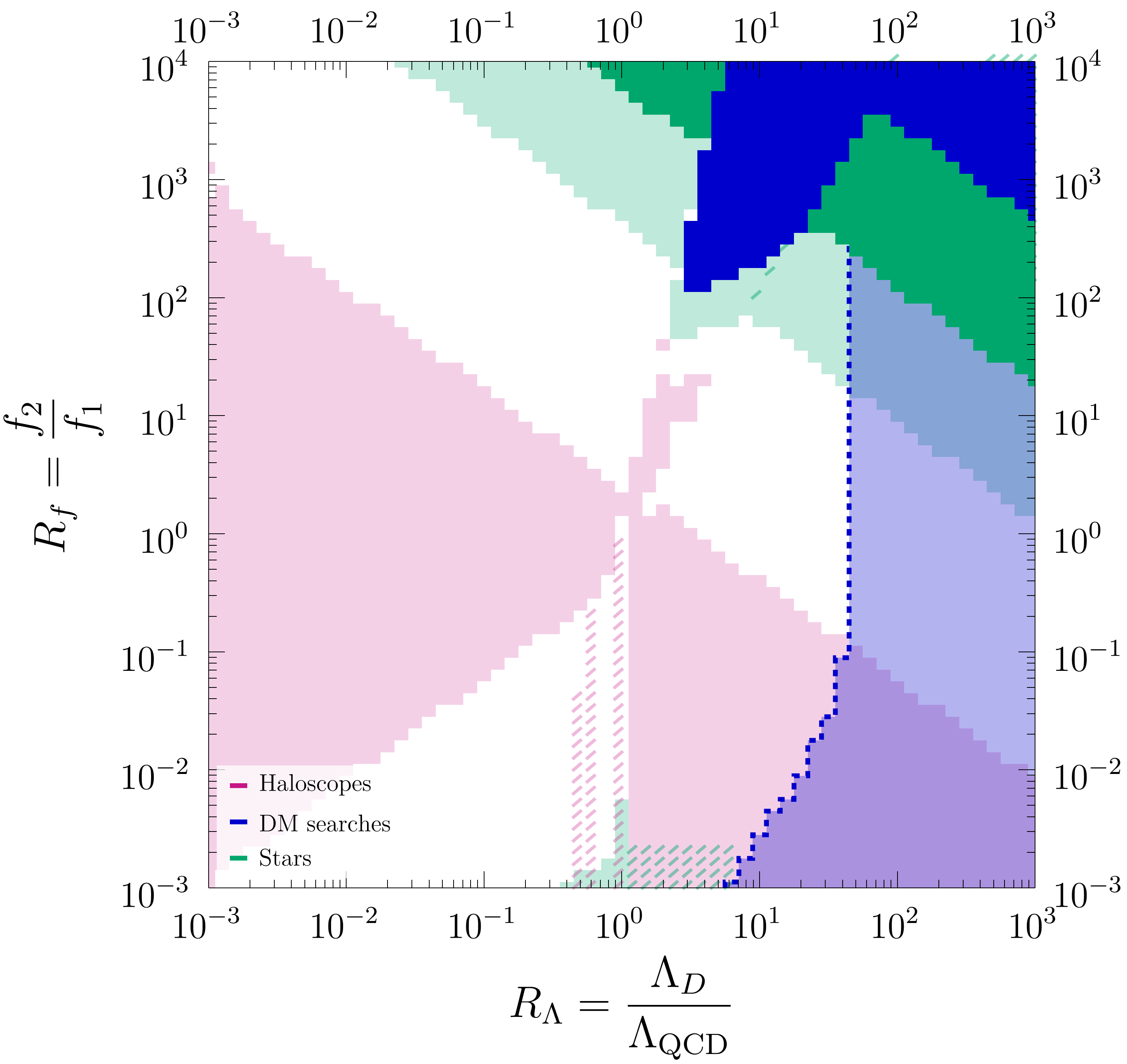}\hfill
  \includegraphics[width=0.48\textwidth]{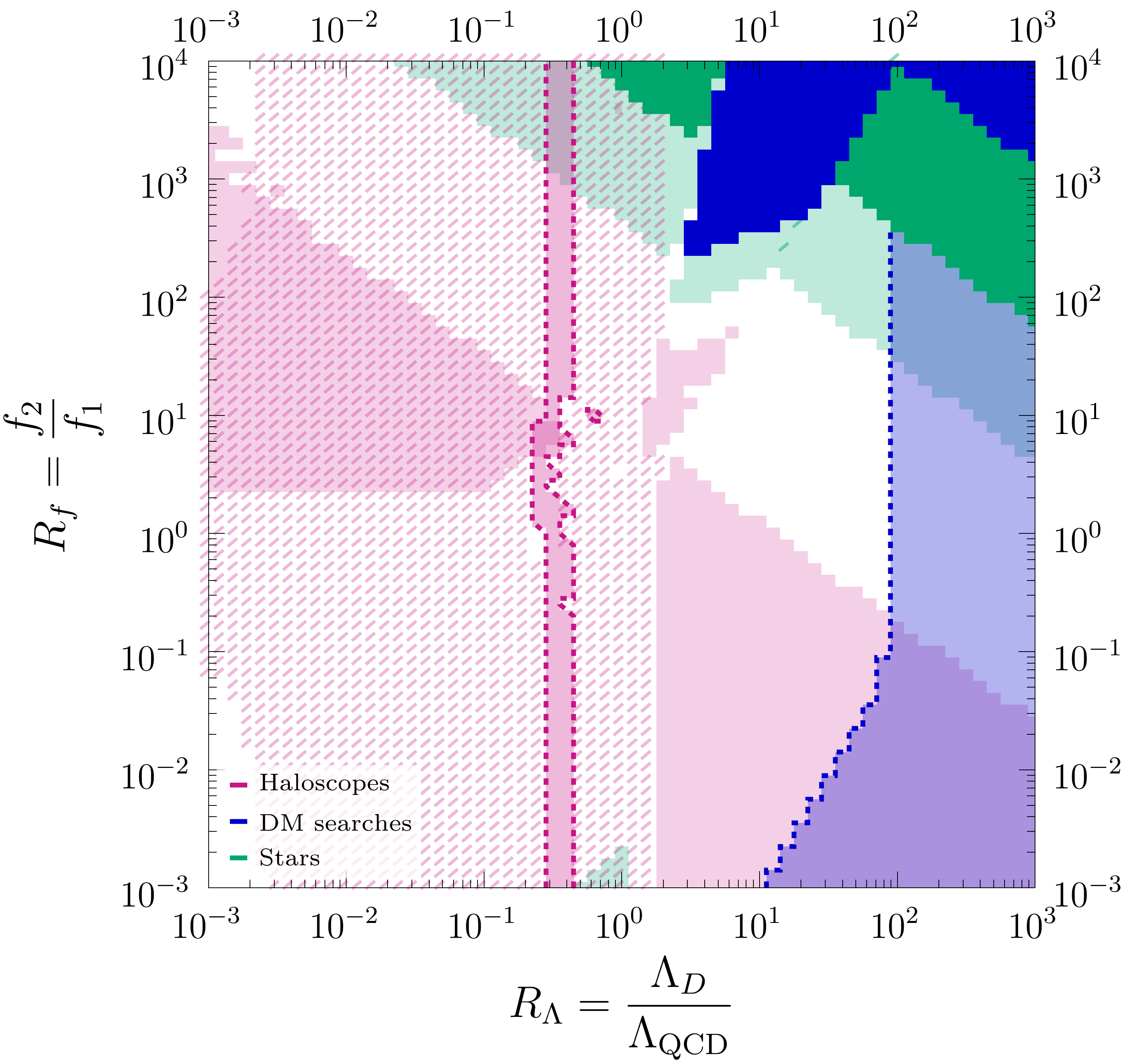}
  \caption{Constraints on the minimal QCD axion photon coupling for different scenarios, assuming that the model explains the totality of dark matter observed. In the left panels, we set $j_2=1$, $k_1 = k_2 = 4/3$, while in the right panels $j_2=3$, $k_1 = 4/3$ and $k_2 = 3$. For all these choices, the QCD axion coupling is minimal in RegI-III. The upper and lower panels differ in the assumption on the dark axion mass: in the upper (lower) panels, it scales as $(T/T_D)^{-\alpha}$ with $\alpha\approx 0$ ($\alpha\approx n$). The parameter space excluded by current bounds on the QCD (dark) axion is enclosed by solid (dotted) lines, while regions accessible to future experiments are shown in light color (hatched). Different colors indicate the type of probe setting the strongest constraint. Here, ``DM searches" stands for cosmological searches based on the dark matter fraction~\cite{AxionLimits}. In the first panel, we identify the haloscope experiments which are more sensitive to the QCD axion, even though others can also probe partially this parameter space.
  }
    \label{fig:QCD photon coupling}
\end{figure}

\begin{figure}
  \centering
  \includegraphics[width=0.8\textwidth]{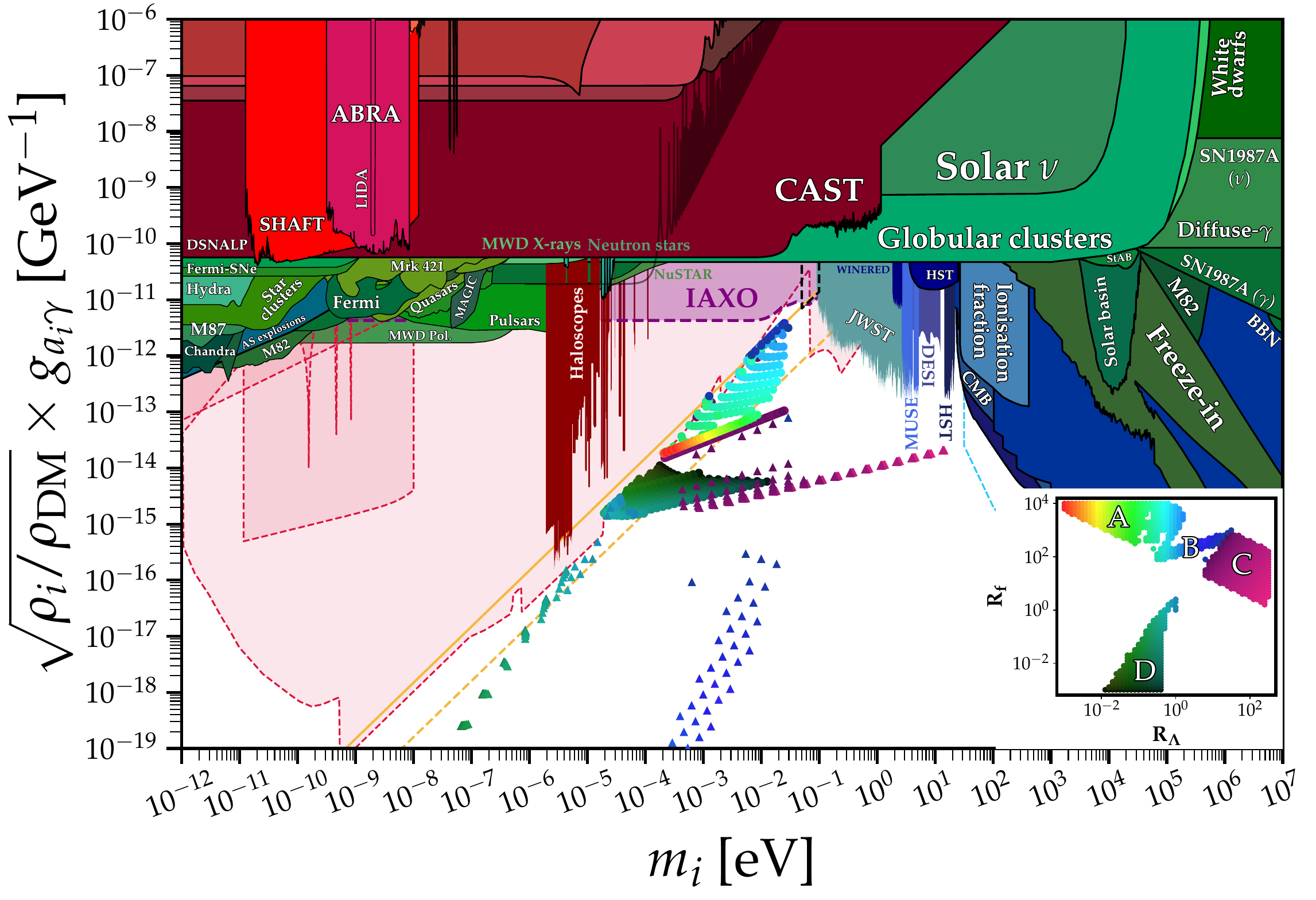}\hfill
  \caption{Representation of QCD (bullet) and dark axion (triangle) pairs that evade future projections, in the scenario depicted in the upper left panel of Fig.~\ref{fig:QCD photon coupling}. The letters refer to the parameter regions defined therein, as shown in the sub-plot.
  The orange solid (dashed) line identifies the $E/N=8/3\,(2)$ benchmark.
  The light red color represents projected sensitivies for several haloscope experiments~\cite{AxionLimits}. We note that helioscope bounds are not rescaled.
  }
    \label{fig:photon coupling vs mass for equal k}
\end{figure}

\begin{figure}
  \centering
  \includegraphics[width=0.8\textwidth]{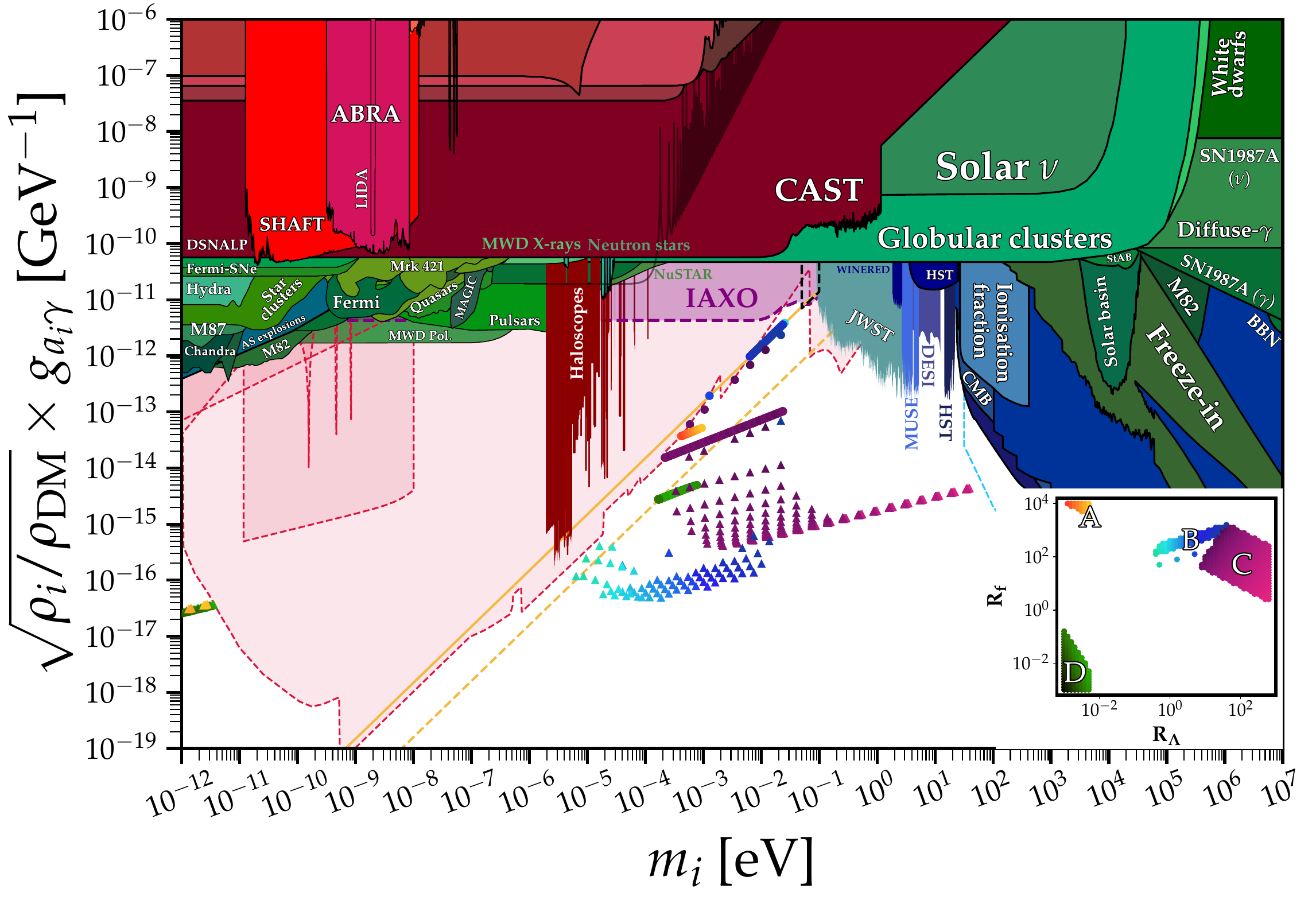}\hfill
  \caption{Representation of QCD (bullet) and dark axion (triangle) pairs that evade future projections, in the scenario depicted in the upper right panel of Fig.~\ref{fig:QCD photon coupling}. All other labels are as in Fig.~\ref{fig:photon coupling vs mass for equal k}.}
    \label{fig:photon coupling vs mass}
\end{figure}

Given the analyses of the previous sections, we have re-interpreted current and future constraints on the two axion model. This will allow us to propose further extensions of the experimental programme to probe motivated regimes of this multi-axion scenario.

The first results are shown in Fig.~\ref{fig:QCD photon coupling}, where 
we represent combined bounds on ``nightmare" scenarios for a post-inflationary QCD axion, assuming the minimal photon coupling allowed by the minimal SM gauge group. The precise values of the photon couplings are plotted in App.~\ref{app:Axion-Photon_Couplings}.
At each point of the parameter space, the model is free from a DW problem and the two axions explain the totality of dark matter observed. 

The assumptions that entail the four panels in this figure are the following.
In the upper panels, we assume a dark axion mass which is roughly $T$-independent, whereas in the bottom panels the dark axion mass follows the same power law as the QCD axion mass becoming constant at $T\approx \Lambda_D$.
The left and right panels showcase different choices for the anomalous couplings. On the left, we set $j_2 = 1$ as well as $k_1 = k_2 = 4/3$. Therefore, the minimal QCD axion coupling arises for $E/N=8/3$ across the parameter space.\,\footnote{This is true except in tuned regions e.g. $R_f\sim 1$ or $R_{\Lambda}\sim 1$; see Fig.~\ref{fig:regimes}.} On the right, $j_2 = 3$ while $k_1  = 4/3$ and $k_2 = 3$. The minimal QCD axion coupling is then the same as in the left panels except for RegIII ($R_f\,,R_\Lambda < 1$), where it occurs for $E/N=2$; see Tab.~\ref{tab:quantization}.

The figure shows that regions where the QCD axion dominates can be largely probed by next-generation experiments.
This is expected in regimes where the single axion model predictions are recovered, for instance in RegI and RegIII with $\rho_{\rm QCD}/\rho_D > 1$.
RegII, where predictions are modified by level crossing, can be partially probed by next-generation helioscopes, such as IAXO$+$~\cite{Armengaud_2019}, displayed in light green in the figure.

The results also demonstrate that there are sizable regions of parameter space where the QCD axion eludes future searches, e.g. in RegIII of the right panels. This holds even within a minimal gauge group setting, which sets a lower bound on the photon coupling to avoid the DW problem. Nevertheless, in these regions, the dark axion becomes a promising signal to probe the setup at next-generation haloscope experiments, such as ADMX~\cite{stern2017admxstatus}. 
Moreover, current astrophysical bounds based on the cosmic background~{\cite{Porras-Bedmar:2024uql} are sensitive to the dark axion coupling in the $R_\Lambda \gg 1$ region (colored in blue).
Altogether, these constraints shed a new light on discovery prospects of multi-axion scenarios where the solutions to the strong CP problem and dark matter are shared between distinct fields.

However, the detectability of the dark axion depends on its anomaly coefficients. While the right panels of Fig.~\ref{fig:photon coupling vs mass} show scenarios where the dark axion is within reach of future experiments, the left panels correspond to scenarios where a GUT symmetry renders its photon coupling vanishing at leading order, causing it to evade detection; see Eqs.~\eqref{eq:CregII} and~\eqref{eq:CregIII}.

We therefore turn to the regions of parameter space evading future experimental reach, labeled A–D in the upper left panel of Fig.~\ref{fig:QCD photon coupling}, and show the location of the two axions simultaneously explaining dark matter and the strong CP problem in Fig.~\ref{fig:photon coupling vs mass for equal k}. Each region gives rise to several paired signals: the colored circle indicates the QCD axion position, while the corresponding triangle marks that of the dark axion.

Most of the QCD axion signals in the figure correspond to values of $E/N$ in the range $[8/3,\, 2]$. 
An order of magnitude improvement beyond the projected sensitivity of haloscope experiments could potentially probe most of these signals.
Interestingly, we find a lower bound on the QCD axion photon coupling of around $10^{-15}\,\text{GeV}^{-1}$, corresponding to points in region D. 
This arises because, even though the dark axion is dominant, its coupling is constrained by astrophysical and haloscope bounds; since both axion couplings depend on the same potential parameters, this translates into a lower bound on the QCD axion photon coupling, regardless of its dark matter fraction.
Furthermore, in several of these regions the QCD axion mass is larger than in the single axion model, which enhances its photon coupling.

We now discuss in more detail the distinct sets of points in Fig.~\ref{fig:photon coupling vs mass for equal k}.
The QCD axions in region A are subdominant, but their coupling to photons is close to the projected IAXO+ limit (helioscope bounds are not rescaled in the plot due to the different dark matter fractions spanned by the points).
In this region, the dark matter abundance is explained by the dark axion, which misaligns with a QCD $T$-dependent mass; see Fig.~\ref{fig:regimes}. 
However, due to level crossing, its relic abundance is set by a combination of $f_1$ and $f_2$ which allows the dark axion to develop a very small coupling to photons, set by $f_2$, while explaining dark matter.
This, together with the {leading order} cancellation in the coupling in Eq.~\eqref{eq:CregII}, pushes the dark axions beyond the parameter space represented in the figure.

In region B instead, the QCD axions are the dominant dark matter component (so they are located in the $E/N =8/3$ line, corresponding to the minimal coupling to photons in our setup), but they have an enhanced mass due to level crossing.
The potential signals from this region populate an interesting region of parameter space between the projected sensitivities of BREAD~\cite{Liu_2022} and LAMPOST~\cite{Baryakhtar_2018}, currently not covered by any planned experiment.

In region C, the dark axions represent the dominant dark matter component.
The line of points observed here is the typical axion-like particle misaligment line, obtained for a $T$-independent mass. 
The dominance of the dark axion is maintained in region D.
However, in this case, they misalign with a QCD $T$-dependent mass.
Due to the absence of level crossing in this region, 
the dark axions populate a parameter space closer to experimental sensitivities in comparison to region A.

In the transition regions between B, C and D, the absence of signals in the plot indicates that one of the axions becomes testable at next-generation experiments, 
such as ADMX~\cite{Fan_2025} and BREAD~\cite{Liu_2022}; see Fig.~\ref{fig:QCD photon coupling}.

We also display the axion pair locations for the most elusive regions of the non-GUT scenario shown in the upper right panel of Fig.~\ref{fig:QCD photon coupling}. The interpretation of 
 Fig.~\ref{fig:photon coupling vs mass} is entirely analogous to that of Fig.~\ref{fig:photon coupling vs mass for equal k}, with the exception that dark axions can develop larger photon couplings, even above the QCD axion line. This motivates searches in this region beyond future projections, and accounts for why most of RegIII is within reach of future haloscopes in this scenario.

\section{Conclusions}\label{sec:conclusions}

In this work, we have shown that theoretical arguments based on minimality often fail.
The multiplicity of axions in the low-energy theory of Nature can significantly impact the properties of the QCD axion, even if the additional axion-like fields decouple from the IR phenomenology, namely the solution to the strong CP problem.

Assuming a post-inflationary scenario, we have quantified the combined effects of novel emergent phenomena in multi-axion theories, including level crossings and the formation of string bundles.
We have shown that these affect not only the cosmology of the QCD axion and its relic abundance, but also the theoretical bounds on its parameter space that follow from the minimal Standard Model gauge group.
The reason is that a key assumption underlying such bounds -- the absence of long-lived domain walls -- is governed by the full anomaly structure of the multi-axion theory.

Working in an illustrative two-axion setup, we found that the interplay with a dark axion can drive the QCD axion towards invisibility: not only does $ E/N < 8/3 $ become compatible with the minimal gauge group setting, but the dark matter fraction carried by the QCD axion also decreases.
Yet in most of this parameter space, the dark axion emerges as a promising signal for next-generation haloscope experiments, a feature we expect to hold beyond the two-axion setup provided the model explains the totality of dark matter.

Nevertheless, cancellations in the dark axion coupling can push it further from the QCD axion band, where most experiments are concentrated.
Such cancellations can be enforced by a GUT symmetry relating the anomalous couplings of each axion to the Standard Model gauge fields.
Even in this case, predictions for the signal regions of the subdominant QCD axion remain possible: we find that its parameter space is bounded from below by
${g_{a\gamma}\gtrsim 10^{-15}\,\text{GeV}^{-1}}$ and $m_a \gtrsim 10^{-5}$\,eV.
No experimental projections are currently expected to cover these regions.

Interestingly, some regions where the QCD axion constitutes the dominant dark matter component also lie beyond future experimental projections.
This is a consequence of level crossing in the early universe, which causes the QCD axion to misalign with the dark axion decay constant, developing a larger mass of around $10^{-2}$\,eV while fully accounting for the observed relic abundance.
Such points cluster at or above the $E/N=8/3$ line near the projected BREAD limit, defining a novel and motivated region of QCD axion parameter space.
Covering these regions would require either new haloscope experiments or improved helioscope sensitivity, with IAXO+ already probing some of these points.

Overall, our analysis demonstrates that, even in ``nightmare'' scenarios for QCD axion detection, other axion-like fields generically emerge as promising signals for next-generation experiments. 
Moreover, even when the QCD axion is under-abundant, its properties are correlated to those of the dominant field, placing a lower bound on its parameter space.
Taken together, these results establish that the multi-axion framework remains testable across most of the parameter space.

\section*{\textbf{Acknowledgments}}

We thank Sungwoo Hong, Mario Reig and Fuminobu Takahashi for insightful discussions. We also thank Guilherme Guedes for valuable feedback on the manuscript. M.R. and F.V. acknowledge support from the COST Action ''Cosmic WISPers in the Dark Universe: Theory, astrophysics,
and experiments” (CA21106). F.V. thanks very much the CERN-TH group for their warm hospitality, where part of this work was carried out. F.V. also acknowledges support from MCIN/AEI (10.13039/501100011033) and ERDF through grants PID2022-139466NB-C21 and PID2022-139466NB-C22; from the Junta de Andalucía through grant P21-00199; from the Consejería de Universidad, Investigación e Innovación, Gobierno de España, and Unión Europea – NextGenerationEU through grant AST22 6.5; and from the Junta de Andalucía / CUII and FSE+ through grant DGP\_PRED\_2024\_01613.

\bibliography{axion}

\clearpage

\appendix

\section{Magnitude of Axion Photon Couplings}
\label{app:Axion-Photon_Couplings}

Figures~\ref{fig:qcd photon coup} and~\ref{fig:dark photon coup} show the QCD and dark photon couplings for the scenarios represented in Fig.~\ref{fig:QCD photon coupling}. In the upper panels, we assume a dark axion mass which is roughly $T$-independent, whereas in the bottom panels the dark axion mass follows the same power law as the QCD axion mass.
The left and right panels showcase different choices for the anomalous couplings. On the left, we set $j_2 = 1$ as well as $k_1 = k_2 = 4/3$, while in the right panels $j_2=3$, $k_1 = 4/3$ and $k_2 = 3$.

\begin{figure}[h]
  \centering
  \includegraphics[width=0.48\textwidth]{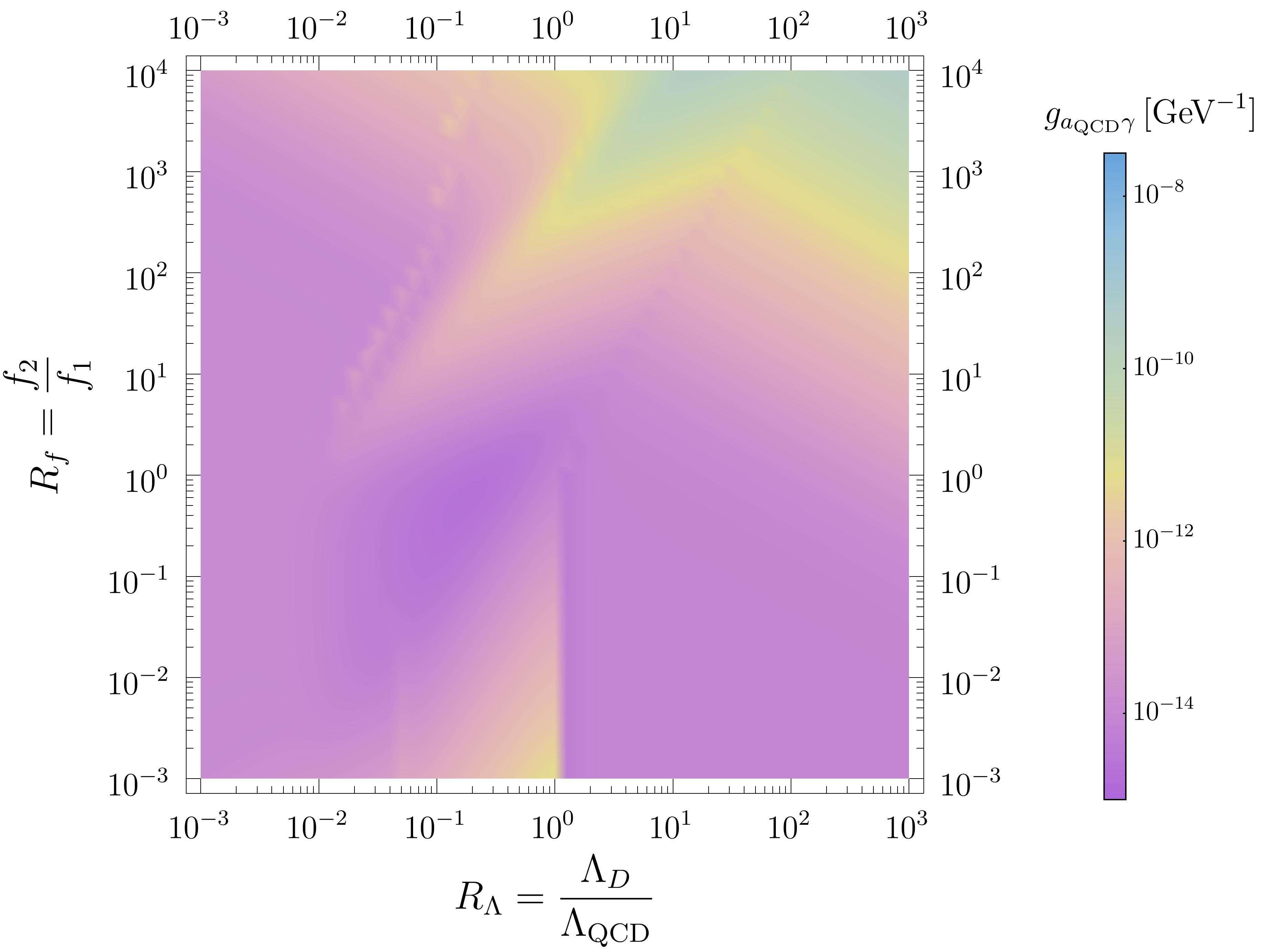}
  \includegraphics[width=0.48\textwidth]{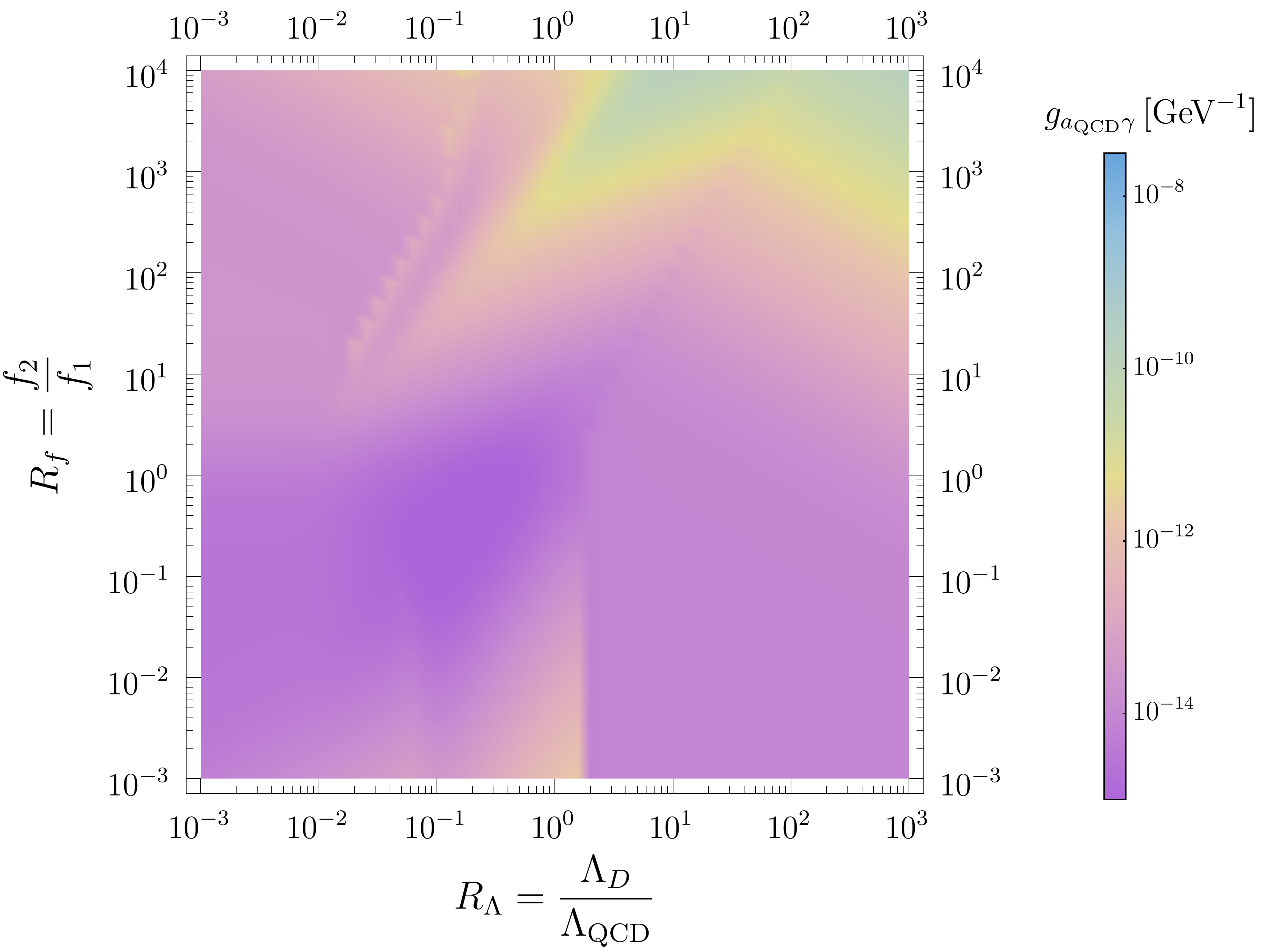}
  \includegraphics[width=0.48\textwidth]{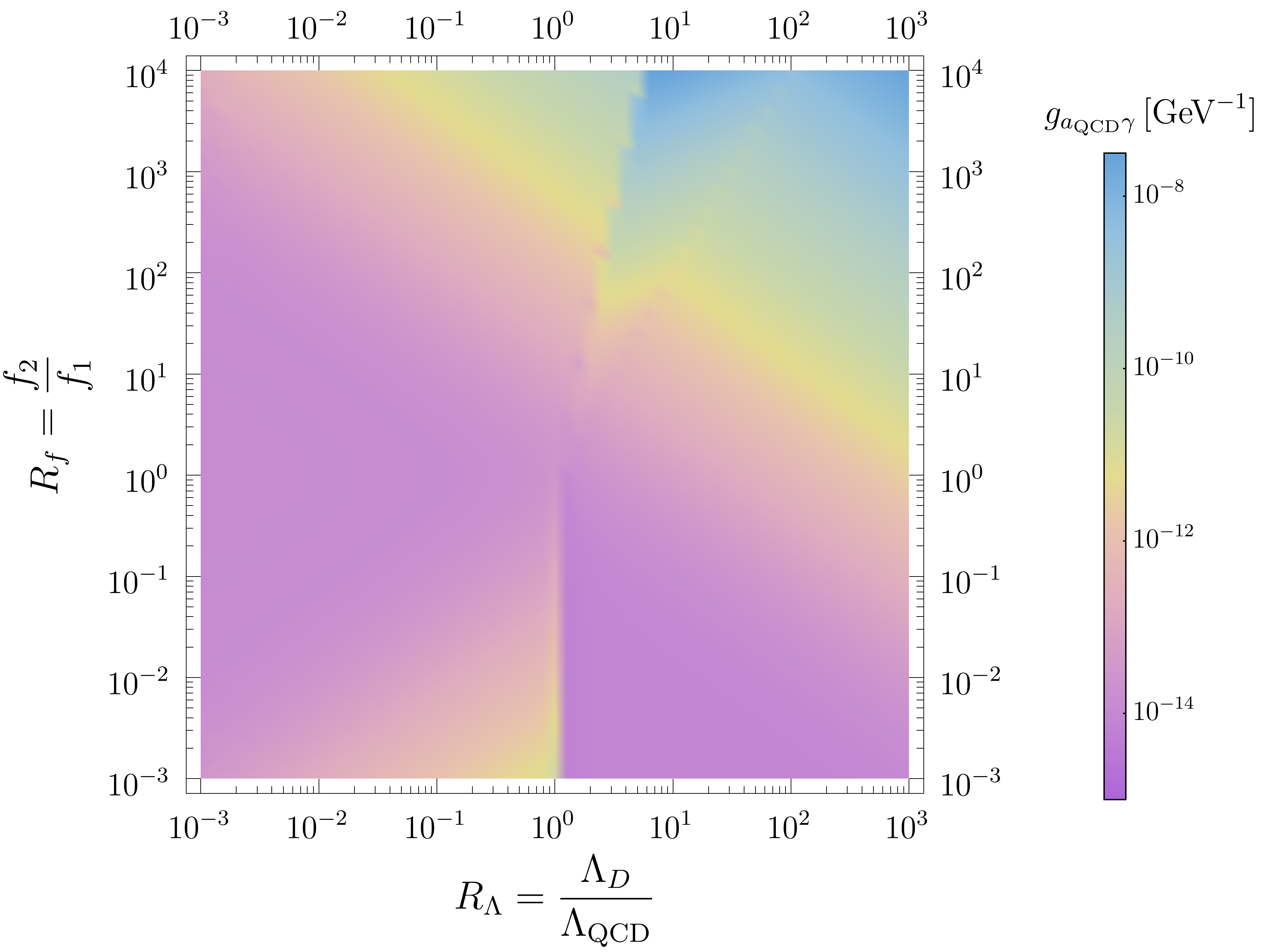}
  \includegraphics[width=0.48\textwidth]{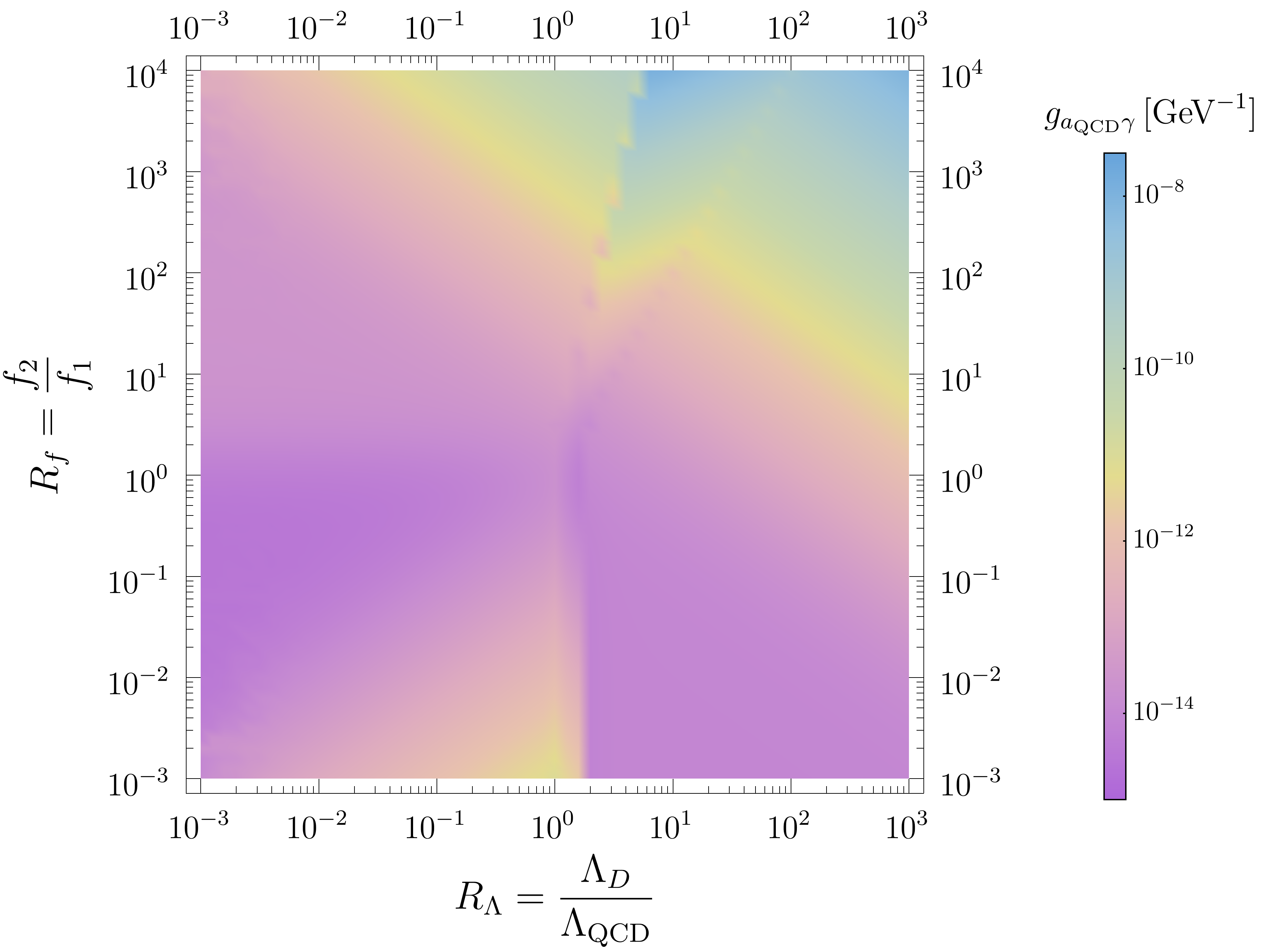}
  \caption{The variation of the QCD axion photon coupling across the parameter space in the scenarios represented in Fig.~\ref{fig:QCD photon coupling}; see the text for details.}
    \label{fig:qcd photon coup}
\end{figure}

\begin{figure}[h]
  \centering
  \includegraphics[width=0.48\textwidth]{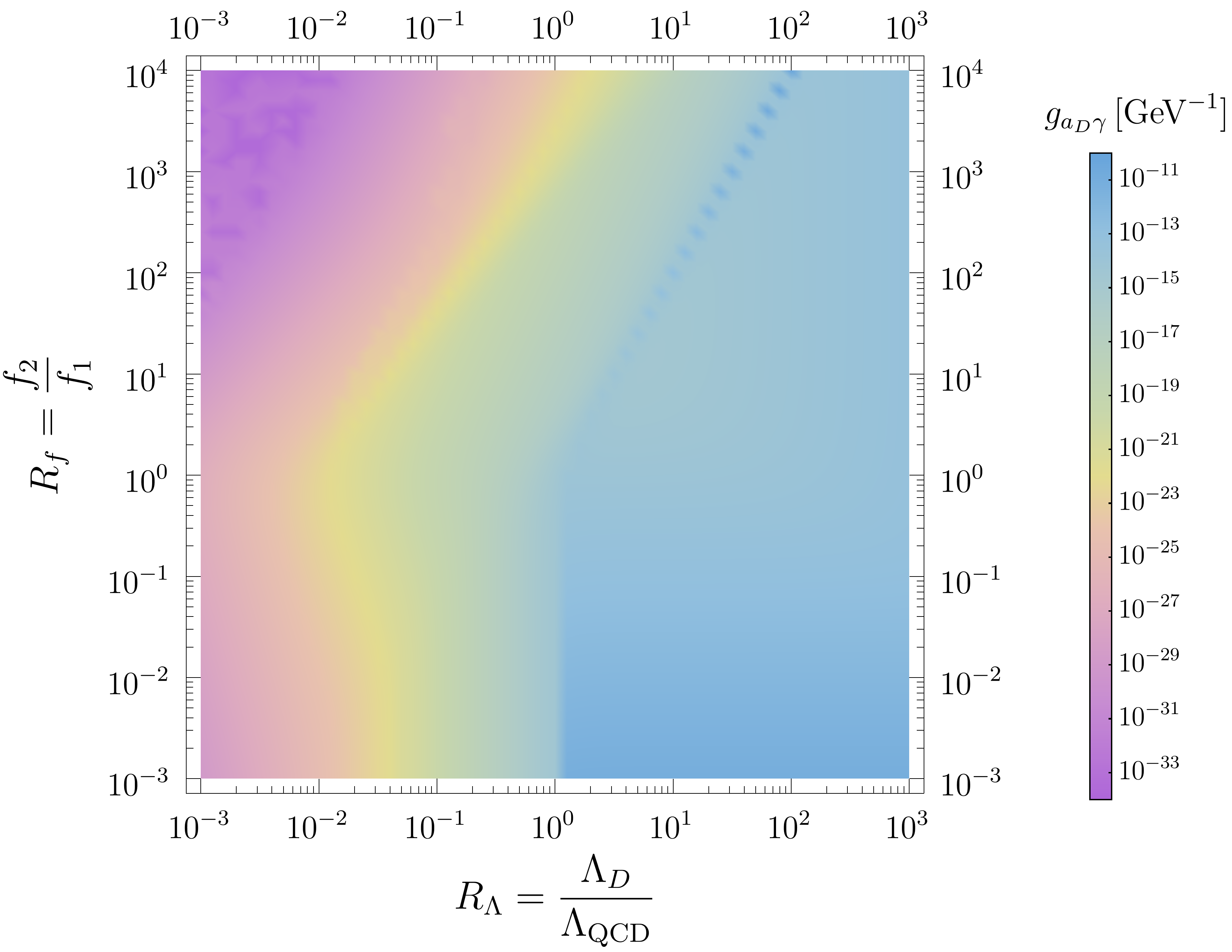}
  \includegraphics[width=0.48\textwidth]{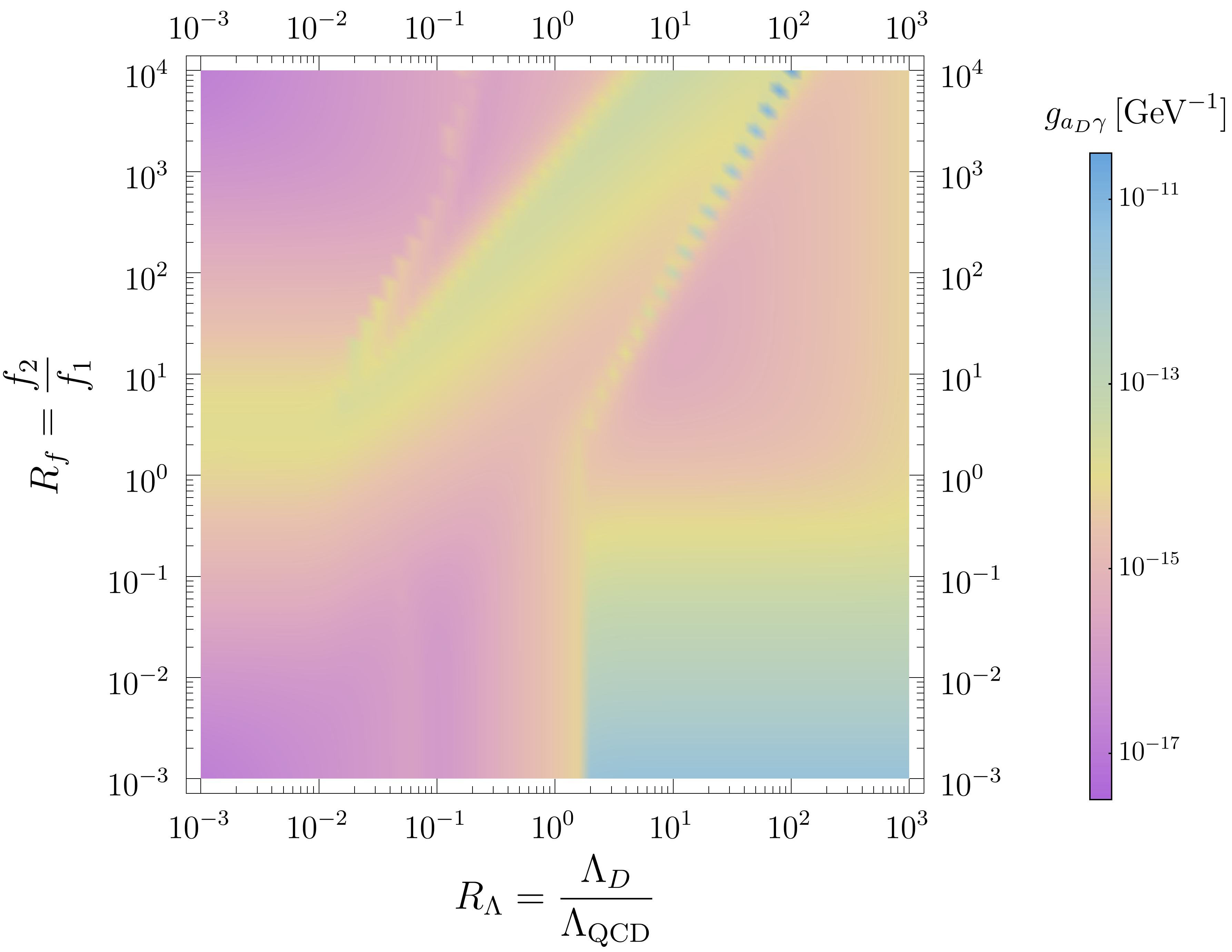}
  \includegraphics[width=0.48\textwidth]{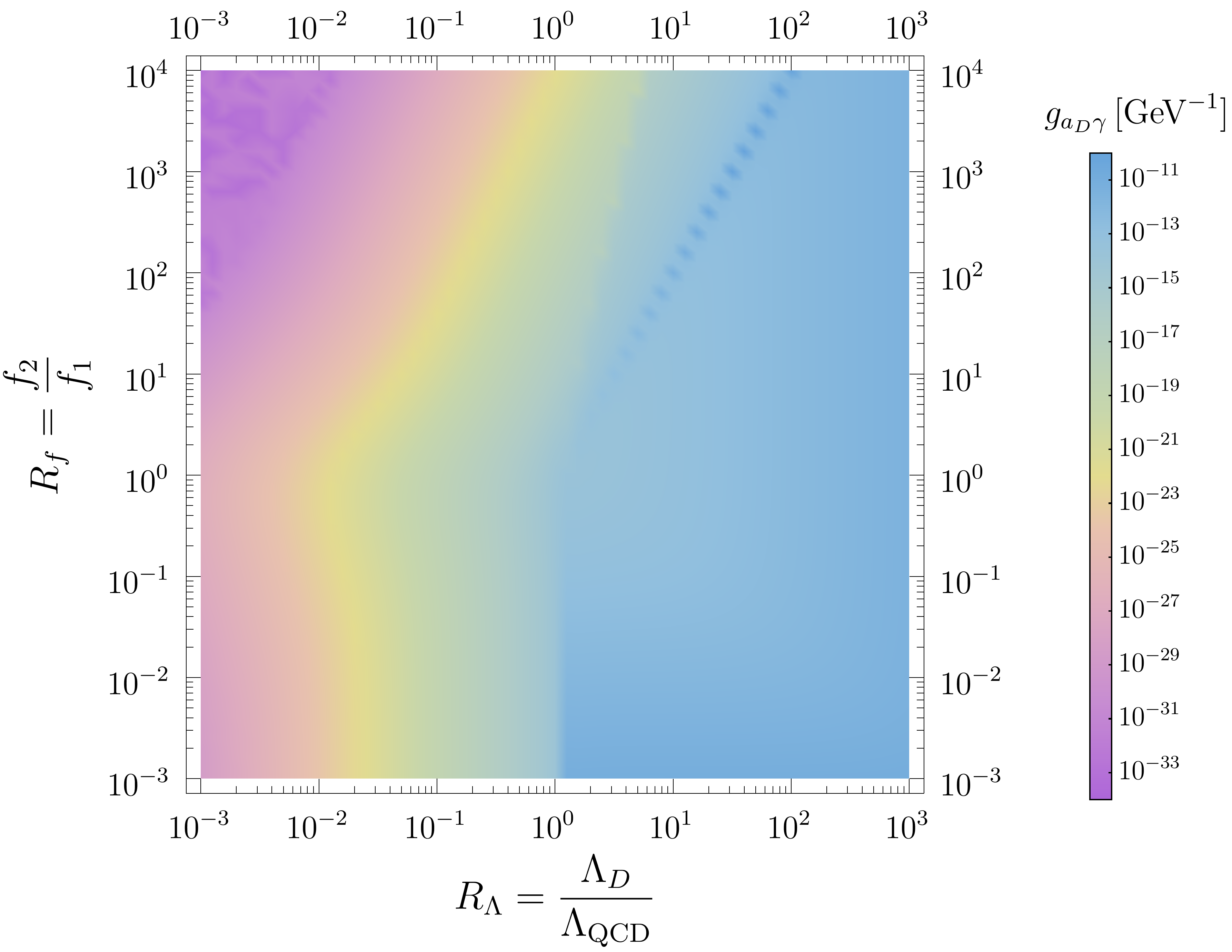}
  \includegraphics[width=0.48\textwidth]{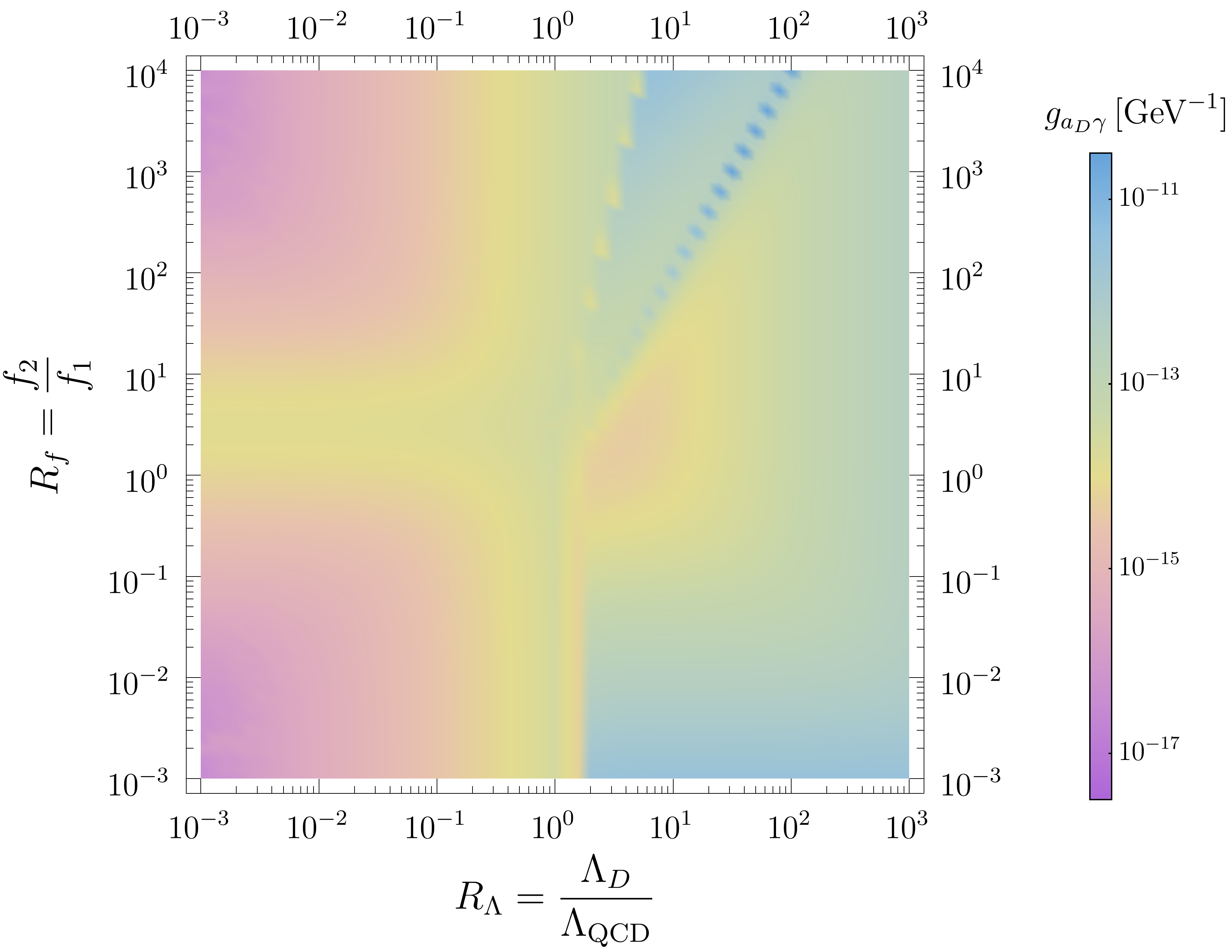}
  \caption{Same as Fig.~\ref{fig:qcd photon coup} but for the dark axion photon coupling.}
    \label{fig:dark photon coup}
\end{figure}

\end{document}